\documentclass[journal]{IEEEtran}
\usepackage{fix-cm} 
\usepackage{tracefnt} 
\usepackage{cite}
\usepackage{amsmath,amssymb,amsfonts}
\usepackage{algorithm}
\usepackage{graphicx}
\usepackage{textcomp}
\usepackage{stfloats}
\usepackage{xcolor}
\usepackage{glossaries}
\usepackage{siunitx}
\usepackage[caption=false]{subfig}
\usepackage[noend]{algpseudocode}
\usepackage{gensymb} 
\usepackage{booktabs} 
\usepackage{xfrac} 
\usepackage{cleveref}
\usepackage{mathtools} 
\usepackage{afterpage}
\usepackage{capt-of}

\usepackage[absolute]{textpos}
\graphicspath{{images/}}

\def\BibTeX{{\rm B\kern-.05em{\sc i\kern-.025em b}\kern-.08em
    T\kern-.1667em\lower.7ex\hbox{E}\kern-.125emX}}

\usepackage{balance}

\newacronym{VR}{VR}{Virtual Reality}
\newacronym[plural=HMDs]{HMD}{HMD}{Head-Mounted Display}
\newacronym{mmWave}{mmWave}{millimeter-wave}
\newacronym{AP}{AP}{Access Point}
\newacronym{(MU-)MIMO}{(MU-)MIMO}{(Multi-User) Multiple Input Multiple Output}
\newacronym{ULA}{ULA}{Uniform Linear Array}
\newacronym{URA}{URA}{Uniform Rectangular Array}
\newacronym{Slerp}{Slerp}{Spherical linear interpolation}
\newacronym[plural=UAVs, longplural=Unmanned Aerial Vehicles]{UAV}{UAV}{Unmanned Aerial Vehicles}
\newacronym{COTS}{COTS}{Commercial Off-The Shelf}
\newacronym[plural=AoAs]{AoA}{AoA}{Angle of Arrival}
\newacronym{SNR}{SNR}{Signal-to-Noise Ratio}
\newacronym{LoS}{LoS}{Line-of-Sight}
\newacronym{EIRP}{EIRP}{Effective Isotropic Radiated Power}
\newacronym{AWV}{AWV}{Antenna Weight Vector}
\newacronym{MCS}{MCS}{Modulation and Coding Scheme}
\newacronym{V2X}{V2X}{Vehicle-To-Everything}
\newacronym{HRLLC}{HR2LLC}{High Rate, High Reliability and Low Latency Communications}
\newacronym{QoE}{QoE}{Quality of Experience}
\newacronym{NLoS}{NLoS}{Non Line-of-Sight}
\newacronym[plural=MPCs]{MPC}{MPC}{Multi-Path Component}
\newacronym[plural=AoDs]{AoD}{AoD}{Angle of Departure}
\newacronym{CDF}{CDF}{Cumulative Distribution Function}
\newacronym{6DoF}{6DoF}{Six Degrees of Freedom}
\newacronym{UE}{UE}{User Equipment}

\DeclareSIUnit{\belmilliwatt}{Bm}
\DeclareSIUnit{\belisotropic}{Bi}
\DeclareSIUnit{\dBm}{\deci\belmilliwatt}
\DeclareSIUnit{\dBi}{\deci\belisotropic}
\DeclareSIUnit{\Gbps}{Gbps}
\DeclareSIUnit{\rad}{rad}
\DeclareSIUnit{\uv}{uv}

\newcommand*{\compconj}{^{\mathrm{*}}}
\newcommand{\oneraggedpage}{\let\mytextbottom\@textbottom
  \let\mytexttop\@texttop
  \raggedbottom
  \afterpage{%
  \global\let\@textbottom\mytextbottom
  \global\let\@texttop\mytexttop}}

\MakeRobust{\Call} 
\algnewcommand\algorithmicto{..}
\algrenewtext{For}[3]{\algorithmicfor\ #1 $\gets$ #2\algorithmicto#3 \algorithmicdo}

\begin{document}

\title{CoVRage: Millimeter-Wave Beamforming for Mobile Interactive Virtual Reality}
\author{Jakob Struye \IEEEmembership{Student Member, IEEE}, Filip Lemic \IEEEmembership{Member, IEEE} and Jeroen Famaey \IEEEmembership{Senior Member, IEEE}
\thanks{J. Struye and J. Famaey are with IDLab - Department of Computer Science, University of Antwerp - imec, Antwerp, Belgium (email: firstname.lastname@uantwerpen.be)}
\thanks{F. Lemic is with i2Cat Foundation, Spain and University of Antwerp - imec, Belgium (email: filip.lemic@i2cat.net)}
}
\maketitle
\begin{abstract}
Contemporary Virtual Reality (VR) setups often include an external source delivering content to a Head-Mounted Display (HMD). ``Cutting the wire'' in such setups and going truly wireless will require a wireless network capable of delivering enormous amounts of video data at an extremely low latency. The massive bandwidth of higher frequencies, such as the millimeter-wave (mmWave) band, can meet these requirements. Due to high attenuation and path loss in the mmWave frequencies, beamforming is essential. In wireless VR, where the antenna is integrated into the HMD, any head rotation also changes the antenna's orientation. As such, beamforming must adapt, in real-time, to the user's head rotations. An HMD's built-in sensors providing accurate orientation estimates may facilitate such rapid beamforming. In this work, we present coVRage, a receive-side beamforming solution tailored for VR HMDs. Using built-in orientation prediction present on modern HMDs, the algorithm estimates how the Angle of Arrival (AoA) at the HMD will change in the near future, and covers this AoA trajectory with a dynamically shaped oblong beam, synthesized using sub-arrays. We show that this solution can cover these trajectories with consistently high gain, even in light of temporally or spatially inaccurate orientational data.
\end{abstract}


\section{Introduction}
\begin{textblock}{180}(20,265)
\begin{tiny}
\hspace{-3.5mm}Accepted to IEEE Transactions on Wireless Communications\\
\vspace{-6mm}\\
\copyright\ 2022 IEEE. Personal use of this material is permitted. Permission from IEEE must be obtained for all other uses,\\
\vspace{-6mm}\\
in any current or future media, including reprinting/republishing this material for advertising or promotional purposes,\\
\vspace{-6mm}\\
creating new collective works, for resale or redistribution to servers or lists, or reuse of any copyrighted component\\
\vspace{-6mm}\\
of this work in other works
\end{tiny}
\end{textblock}
\IEEEPARstart{M}{obile}, interactive and collaborative \gls{VR}, where a wireless connection provides a \gls{HMD} with an audiovisual stream recorded or generated in real-time, requires a reliable low-latency connection with massive throughput~\cite{VRChallenges}. While some \glspl{HMD} can generate content on-device, applications requiring high-fidelity, high-framerate, and high-resolution content call for a connection to a powerful machine. Current devices, such as the stand-alone Meta Quest 2, provide this capability through a wired connection or a \SI{5}{\giga\hertz} wireless home network. The former restricts the player's freedom of movement and poses a tripping hazard, hindering immersion~\cite{VRChallenges}, while the latter induces noticeable video compression.

To achieve truly wireless connected \glspl{HMD}, \gls{mmWave} networking is most often considered. Operating at frequencies of \SI{30}{} to \SI{300}{\giga\hertz}, \gls{mmWave} is capable of meeting the \gls{VR} requirements~\cite{VRChallenges}. Solutions often rely on the existing IEEE 802.11ad/ay Wi-Fi standards for \gls{mmWave}~\cite{PerasoVR,cotsMMVR} or on 5G NR's \gls{mmWave} capabilities~\cite{VrMecFallback}.
The main challenges in building such a system stem from \gls{mmWave}'s inherently high path loss and attenuation. To achieve sufficiently high signal strength at the \gls{HMD}, the transmitter and the \gls{HMD} must both focus their energy towards each other, both when sending and when receiving, in a process called beamforming. \Gls{mmWave} transceivers usually implement beamforming using phased antenna arrays, consisting of many separate, individually phase-controllable antenna elements~\cite{Fundamentals}. By carefully tuning each element's phase shift, all elements' signals become phase-aligned, and interfere constructively, in some intended direction. 3D beamforming, with two degrees of freedom, requires a 2D grid of antenna elements, such as a \gls{URA}. Beamforming is generally highly challenging, as devices are unaware of other devices' relative positions, often requiring some sort of search algorithm~\cite{compressive,OScan,VirtualHierarchical}. In a 5G context, pose estimations from the UE have been used to facilitate initial access~\cite{5Ginitialaccess} and digital beamforming~\cite{5GDBF}. For \glspl{HMD} specifically, accurate on-device pose estimation~\cite{RotationPrediction} enables direct beamforming between the mobile \gls{HMD} and static \gls{AP}. Several works have leveraged this pose information for beamforming~\cite{MoVR,OScanOLD,OScan,Pia}, noting that \gls{HMD}-side beamforming is considerably more challenging than at the \gls{AP} side. An angular beam misalignment of a few degrees can have a significant impact on \gls{SNR}~\cite{MoVR}, which is most rapidly triggered by head rotations~\cite{OScan}. All of the solutions only generate a beam in a single direction, based on a current pose estimate. During rapid and sudden movement, these algorithms may not be able to update their beam quickly enough. In addition, pose estimates may not be entirely accurate during such motion. As such, we pose that, to maintain mobile \gls{VR}'s network requirements even under rapid motion, \gls{HMD}-side receive beamforming should proactively synthesize an oblong beam of dynamic width expected to continuously cover the \gls{AP} during motion. In this work, we present \textit{coVRage}, the first pose-aware receive-side beamforming algorithm to cover both the current orientation and the predicted trajectory of near-future orientations.

To enable rapid synthesis of such beams, we use sub-arrays. By dividing the full antenna array into separate sub-arrays, each sub-array can form its own sub-beam in a distinct direction. Several approaches to forming these sub-arrays exist. 
With physical sub-arrays, each has its own RF chain. On the other hand, virtual sub-arrays allow for dynamically configurable sub-array layouts. This class is further subdivided into analog arrays, with one RF chain for the full array, and digital arrays, with a separate RF chain for each element~\cite{ShimuraInterleaved}. Virtual sub-arrays enable runtime control over the number of sub-arrays and their shape, in turn facilitating hierarchical codebooks~\cite{subarrayCodebook,VirtualHierarchical}, 3D beams of flexible shape~\cite{FlexibleCoverage} and multiplexing/multi-user transmissions~\cite{subarrayMultiplex}. Our algorithm leverages this feature using analog sub-arrays. We avoid fully digital arrays due to their prohibitively high cost and power consumption for a mobile device. 

\begin{figure*}[!t]
    \centering
    \begin{minipage}{1.0\linewidth}
    \subfloat[Localized sub-arrays]{\includegraphics[width=0.27\textwidth]{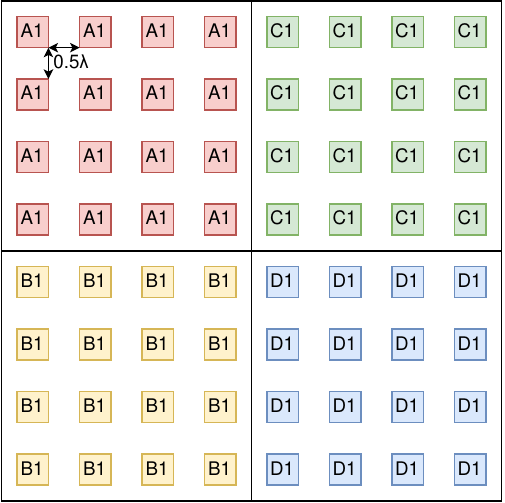}\label{fig:localized}}
    \hfill
    \subfloat[Interleaved sub-arrays]{\includegraphics[width=0.27\textwidth]{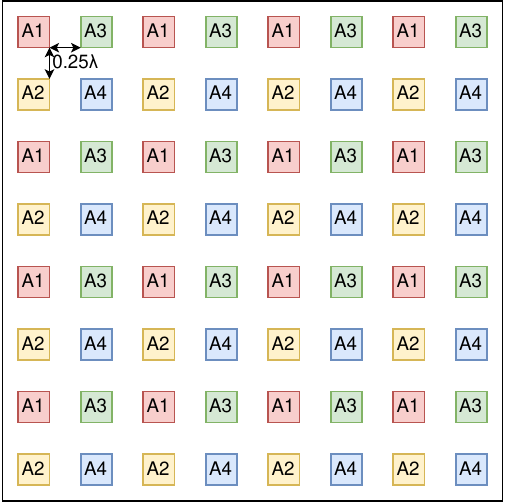}\label{fig:interleaved}}
    \hfill
    \subfloat[Multi-block sub-arrays]{\includegraphics[width=0.27\textwidth]{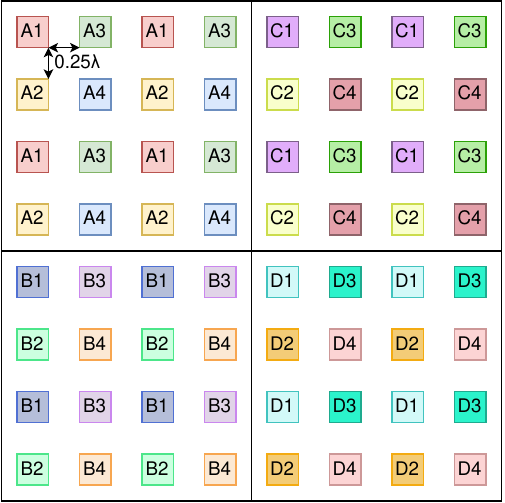} \label{fig:hybrid}}
    \end{minipage}
    \caption{Different (virtual) sub-array layouts in a phased array. A black square designates a block, which is the maximal region within which all sub-arrays are interleaved with each other. Any two sub-arrays from different blocks are localized to each other. The indicated inter-element spacings are chosen to maintain an inter-element spacing of $0.5\lambda$ within a sub-array. Each element label consists of the index of its block (a letter) and the index of its sub-array within that block (a digit).}
    \label{fig:layouts}
\end{figure*}

These sub-arrays, either physical or virtual, can be distributed across the array in several ways. Specifically, a sub-array may be localized, with all elements in the sub-array next to each other, or interleaved, with elements spread across the entire array, illustrated in Fig.~\ref{fig:localized} and \ref{fig:interleaved}~\cite{LocalizedvsInterleaved}. Alternatively, a two-level \textit{multi-block} array consists of localized blocks, each containing interleaved sub-arrays, as in Fig.~\ref{fig:hybrid}. Our algorithm relies on virtual multi-block arrays, which combine features of both localized and interleaved sub-arrays. Interleaved sub-arrays allow for antenna array elements placed closer together physically, utilizing limited \gls{HMD} space more optimally. This reduced spacing may result in adjacent elements influencing each other's signals noticeably, such that mitigation techniques against this \textit{mutual coupling} may be needed. 

To synthesize a beam covering current and future directions towards the \gls{AP}, head rotations must be predicted. Built-in pose estimation present on modern \glspl{HMD}, mainly used for offsetting the expected latency of image generation, may already suffice for this application~\cite{RiftTracking}. Should more accurate prediction be needed, other existing approaches can be applied. For head rotation prediction, many works have shown the effectiveness of classical approaches such as autoregression and Kalman filters~\cite{RotationPrediction,KF5,KF3,KF4}. More recently, the field of viewport prediction, which aims to solve essentially the same problem, has successfully applied more advanced approaches such as graph learning~\cite{viewportGraph}, reinforcement learning~\cite{viewportRL} and recurrent neural networks~\cite{viewportLSTM}. While the different approaches are difficult to compare directly due to varying prediction horizons and datasets, most approaches provide predictions amply accurate for coVRage. In \cite{RotationPrediction}, a relatively straightforward predictor achieves a median error of under \SI{0.35}{\degree} at a \SI{100}{\milli\second} horizon for a dataset with aggressive motion, with a worst case error under \SI{1.7}{\degree}. The dynamic width of our algorithm's beam is therefore not only intended to provide robustness against measurements errors, but also both temporal and spatial errors in prediction.

The main contributions of this paper are summarized as follows:
\begin{enumerate}
    \item We analyze the main challenges and opportunities specific to \gls{mmWave} wireless interactive \gls{VR}. We empirically estimate the expected maximal user motion, showing that rapid receive-side beamforming with orientation prediction is crucial to ensure consistently high \gls{SNR}, which is facilitated by accurate pose measurements gathered by on-device sensors.
    \item We present \textit{coVRage}, a receive-side beamforming algorithm for \glspl{HMD}, supporting uninterrupted connectivity during rapid head movements. This is, to the best of our knowledge, the first \gls{HMD}-focused beamforming method offering proactive \gls{AoA} trajectory coverage through sub-arrays.
    \item  We observe that aiming localized sub-arrays in similar directions causes them to reduce each other's beamwidths, and propose the \textit{transitional sub-array} system to mitigate the effect. Using these, we present several mappings of sub-beams to configurable sub-arrays depending on predicted \gls{AoA} changes and confidence in said predictions.
    \item Using simulation, we demonstrate that coVRage and transitional sub-arrays achieve their design goals, together offering consistently high \gls{SNR} during worst-case head rotations, even when these are imperfectly predicted.
\end{enumerate}

The remainder of this paper is organized as follows. Section~\ref{sec:sysmod} presents the system and channel models used in this work. Section~\ref{sec:probstat} describes the considered problem in detail. Next, Section~\ref{sec:algo} presents the coVRage algorithm, and in Section~\ref{sec:eval} we evaluate how well it performs in simulation. Finally, Section~\ref{sec:conclusion} concludes this paper.

Throughout the article, we use the following notation: $\mathbf{A}$ is a matrix, $\mathbf{a}$ is a vector, $a$ is a scalar. $(\cdot)^\mathrm{T}$ and $(\cdot)^\mathrm{H}$ denote transpose and conjugate transpose. $\angle c$ denotes the angle of complex value $c$. $\mathbf{q}\compconj$ is the complex conjugate (and inverse) of unit quaternion $\mathbf{q}$. $\left\vert\cdot\right\vert$, $\left\Vert\cdot\right\Vert$ and $\left\Vert\cdot\right\Vert_\mathrm{F}$ denote the absolute value, $\ell^2$-norm and Frobenius norm. $\mathbf{I}$ is the identity matrix. $\mathbb{E}\left[\cdot\right]$ denotes the expectation.

\section{System and Channel Model}\label{sec:sysmod}
In this work, we consider a mmWave communication system with, at each end, a square \gls{URA} consisting of $N_T$ and $N_R$ elements at transmit and receive side respectively. As cost and energy efficiency are of prime importance to mobile consumer devices, we assume that each array is driven by a single RF chain, restricting us to analog beamforming. For each element, the phase is individually controllable within the continuous range $[0,2\pi)$. More realistic $n$-bit phase control is modeled by uniformly quantizing phase shifts~\cite{irsQuantize}. Furthermore, each receive element has $1$-bit amplitude control, meaning each element can be turned on or off individually~\cite{VirtualHierarchical}. As the sub-array system is purely virtual, it does not appear in the system model. We present a full system model taking reflections into account, and a simpler \gls{LoS}-only model which is faster to simulate.
\subsection{Full model}
The full system and channel model works as follows. Given a transmitted symbol $s$, with unit power, the received signal becomes
\begin{equation}
    y = \sqrt{P}\mathbf{w}^\mathrm{H}_\mathrm{R}\mathbf{H}\mathbf{w}_\mathrm{T}s + \mathbf{w}^\mathrm{H}_\mathrm{R}\mathbf{n}\label{eq:rcvsig}
\end{equation}
where $P$ is the transmit power, $\mathbf{w}_\mathrm{R} \in \mathbb{C}^{N_R}$ and $\mathbf{w}_\mathrm{T}\in \mathbb{C}^{N_T}$ are the receive and transmit \glspl{AWV} respectively, $\mathbf{H} \in \mathbb{C}^{N_R \times N_T}$ is the channel matrix, and $\mathbf{n}$ is the Gaussian noise vector with power $N_0$, s.t.  $\mathbb{E}\left[{\bf{n}}{\bf{n}}^{\rm{H}}\right]={N_0\bf{I}}$. For an $N$-element square \gls{URA}, the \gls{AWV} is
\begin{equation}
    \mathbf{w} = \nu\left[w_{0,0}, w_{1,0}, \dots, w_{\sqrt{N}-1,\sqrt{N}-1}\right]^\mathrm{T}\label{eq:awv}
\end{equation}
where $\nu$ is a normalization factor s.t. $\left\Vert\mathbf{w}\right\Vert = 1$ and
\begin{equation}
    w_{x,y} = \alpha_{x,y} \exp(j\phi_{x,y}) \label{eq:awv_el}
\end{equation}
where $\alpha_{x,y} \in \{0,1\}$ and $\phi_{x,y} \in [0,2\pi)$ denote amplitude and phase control, respectively.

As channel model, we employ a geometric model, as is commonly used for \gls{mmWave} channels~\cite{geometricModel, VirtualHierarchical, geometric3}. Based on the observation that \gls{mmWave} experiences limited scattering, the model considers a fixed number of paths, or \glspl{MPC}. Given the directional nature of mmWave, each \gls{MPC} is identified by a physical \gls{AoD} ($\theta_d,\psi_d$) and \gls{AoA} ($\theta_a,\psi_a$), with $\theta$ and $\psi$ denoting azimuth and elevation, respectively. The \gls{MPC} gains are Rician-distributed, with one dominant \gls{LoS} \gls{MPC} along with zero or more reflected \gls{NLoS} \glspl{MPC}~\cite{geometricRician}. The channel matrix is then written as
\begin{equation}
    \mathbf{H} = \mu \sum^{L-1}_{l=0}\gamma_l\mathbf{a}(N_R,\theta_{a,l}, \psi_{a,l})\mathbf{a}(N_T, \theta_{d,l}, \psi_{d,l})^\mathrm{H}\label{eq:chmtr}
\end{equation}
where $\mu$ is a normalization factor $\sqrt{N_TN_R}$ s.t. $\mathbb{E}\left[\left\Vert \mathbf{H} \right\Vert_\mathrm{F}^2\right] = N_TN_R$, $L$ is the number of \glspl{MPC}, $\gamma_l$ is the complex coefficient of the $l$-th path, $\mathbf{a}(\cdot)$ is the steering vector, and $(\theta_{d,l}, \psi_{d,l})$ and $(\theta_{a,l}, \psi_{a,l})$ are the \gls{AoD} and \gls{AoA} of the $l$-th path respectively. For every \gls{NLoS} path ($l > 0$), the complex coefficient $\gamma_{nlos}$ is complex Gaussian distributed, and the components of the \gls{AoD} and \gls{AoA} are uniformly distributed in $[0,2\pi)$.
If $L>1$, the K-factor $\mathcal{K}$ controls the \gls{LoS} path's relative power, as
\begin{equation}
    \left\vert\gamma_{los}\right\vert^2 = \mathcal{K}\;(L-1)\;\mathbb{E}\left[\;\left\vert \gamma_{nlos} \right\vert\;\right]^2
\end{equation}
ensuring the \gls{LoS} \gls{MPC}'s power is expected to be $\mathcal{K}$ times as strong as all \gls{NLoS} \glspl{MPC}' combined.
The \glspl{MPC}' complex coefficients are normalized s.t.
\begin{equation}
    \mathbb{E}\left[\sum^{L-1}_{l=0}\left\vert \gamma_l \right\vert ^2\right] = 1
\end{equation}

The steering vector, a function of the array dimensions and the \gls{AoA} or \gls{AoD}, is expressed as
\begin{equation}
    \mathbf{a}(N, \theta, \psi) = \sqrt{\frac{1}{N}}\left[a_{0,0}, a_{1,0}, \dots, a_{\sqrt{N}-1,\sqrt{N}-1}\right]^\mathrm{T} \label{eq:steervec}
\end{equation}
where
\begin{equation}
    a_{x,y} = \exp\left(j\frac{2\pi d}{\lambda}\left(x\sin{\theta}\cos{\psi}+y\sin{\psi}\right)\right)
\end{equation}
where $d$ is the inter-element spacing in both $x$ and $y$-direction and $\lambda$ is the wavelength.

In this system, the beamforming gain is expressed as
\begin{equation}
    G = \left\vert\mathbf{w}^\mathrm{H}_\mathrm{R}\mathbf{H}\mathbf{w}_\mathrm{T}\right\vert^2
\end{equation}

\subsection{Simplified model}
While the system and channel model presented above provides a realistic representation of the actual system, it does imply a practical inconvenience for this work. Specifically, it requires that a physical layout and \gls{AWV} for both receive and transmit antenna arrays be determined, even though our work is a receive-side solution, independent of the transmit side's specifics. We therefore additionally present a simplified, transmit-independent model. For this, we first consider a \gls{LoS}-only version of the full model (i.e., $L=1$). If the transmit beam is aimed perfectly in the \gls{LoS} direction, then $\mathbf{a}(N_T, \theta_{los}, \psi_{los})^\mathrm{H}\mathbf{w}_\mathrm{T} = 1$, with $(\theta_{los}, \psi_{los})$ being the \gls{AoD} of the \gls{LoS} path. In this case, the \gls{AWV} and steering vector can be eliminated from \eqref{eq:rcvsig} and \eqref{eq:chmtr} respectively. When re-introducing \gls{NLoS} paths, the elimination of these factors alters the model's properties, as $\mathbf{a}(N_T, \theta_{d,l}, \psi_{d,l})^\mathrm{H}\mathbf{w}_\mathrm{T} \leq 1$ for each $l^\mathrm{th}$ \gls{NLoS} path ($l > 0$). Eliminating the factors impacts the distribution of \gls{NLoS} path power in two ways: the mean increases, and the variability decreases. Within the model however, these two effects can be offset by increasing $\mathcal{K}$, and increasing the variance of the complex Gaussian distribution $\gamma_{nlos}$ is sampled from, respectively. As such, we argue that the following simplified model still represents the expected environment with sufficient accuracy, while being both faster to compute, and independent of the transmit antenna array's dimensions. The following two equations comprise the simplified model:
\begin{equation}
    y = \sqrt{P}\mathbf{w}^\mathrm{H}_\mathrm{R}\mathbf{H}s + \mathbf{w}^\mathrm{H}_\mathrm{R}\mathbf{n}
\end{equation}
and
\begin{equation}
    \mathbf{H} = \mu \sum^{L-1}_{l=0}\gamma_l\mathbf{a}(N_R,\theta_{a,l}, \psi_{a,l})
\end{equation}
where $\mu$ is now $\sqrt{N_R}$. The effect of a suboptimally aimed transmit beam can still be modeled by increasing $N_0$ and decreasing $\mathcal{K}$.
With this model, the beamforming gain becomes
\begin{equation}
    G = \left\vert\mathbf{w}^\mathrm{H}_\mathrm{R}\mathbf{H}\right\vert^2
\end{equation}
When only considering the \gls{LoS} path (i.e., choosing $\mathcal{K} \rightarrow \infty$, meaning that $\mathbb{E}\left[\;\left\vert \gamma_{nlos} \right\vert\;\right]^2 \rightarrow 0$ and $\left\vert\gamma_{los}\right\vert^2 \rightarrow 1$), we can write this as
\begin{equation}
    G = \left\vert\mu \mathbf{w}^\mathrm{H}_\mathrm{R}\mathbf{a}(N_R,\theta_{los}, \psi_{los})\right\vert^2\label{eq:losgain}
\end{equation}
in which case the beamforming gain equals the receive gain. We consider this single-beam model as sufficiently realistic, as \gls{LoS} is unlikely to be broken for the mobile \gls{VR} use case this work considers. Furthermore, reflections in said use case are most likely via walls, and their power is assumed to be negligible as long as the user is not right next to the wall and grating lobes are avoided. Redirected walking can keep mobile users away from walls in \gls{VR} applications~\cite{RedirectedWalking}.
\section{Beamforming for Mobile Virtual Reality}\label{sec:probstat}
In this section, we outline the challenges and opportunities of \gls{mmWave} \gls{VR}, describe the environment we envision our algorithm to operate in, and present the phased array layouts used for this work.
\subsection{Challenges and Opportunities}
In this work, we consider a \gls{VR} setup where a ceiling-mounted \gls{mmWave} \gls{AP} serves a mobile \gls{VR} user on the ground. The user moves with \gls{6DoF}, and the \gls{HMD} generates high-accuracy estimates of its own pose (i.e., location plus orientation) and can generate short-term pose predictions with reasonable accuracy~\cite{riftAccuracy,RiftTracking}. To ensure a high \gls{QoE}, the downlink from \gls{AP} to \gls{HMD} must have a consistently high \gls{SNR}, even during rapid user movement. From a beamforming perspective, this application is unique on several fronts. In this section, we outline the three main fronts, and discuss the challenges and opportunities they imply for \gls{mmWave} \gls{HMD} beamforming. 

First are the mobility patterns of \glspl{HMD}. For any \glspl{AP} involved, the location and orientation are static and known a priori. \gls{HMD}-wearing users, however, are often encouraged to move around rapidly, especially in truly wireless \gls{VR}. Gauging the impact of both translational (e.g., walking) and rotational (e.g., head rotations) motion using realistic velocities obtained from an experiment shows that rotations are significantly more impactful. Consider an \gls{AP} \SI{2}{\metre} above an \gls{HMD} moving translationally at a walking pace of \SI[per-mode=fraction,fraction-function=\sfrac]{1.4}{\metre\per\second}. In \SI{100}{\milli\second}, this will shift the optimal beam direction by \SI{4}{\degree} for both the \gls{AP} and \gls{HMD}. Instead, a rotational motion of \SI[per-mode=fraction,fraction-function=\sfrac]{300}{\degree\per\second} may, depending on the direction, cause a shift in optimal beam direction of up to \SI{30}{\degree} in \SI{100}{\milli\second} for the \gls{HMD}, with minimal impact on the \gls{AP}. In Sec.~\ref{sec:indiv_traj}, we observe that users can reach a maximum velocity over three times as high in a \SI{100}{\milli\second} window. Clearly, \gls{HMD}-side beamforming under rotational motion is the most challenging sub-problem of mobile \gls{VR} beamforming.

A second front encompasses \gls{VR} network requirements in terms of latency, throughput and reliability. Depending on other sources of latency in the full content delivery chain, the target network latency may be between 1 and \SI{5}{\milli\second}~\cite{VRChallenges}. This implies that, when delivering a live video stream per-image, the interval between subsequent images may be larger than the maximum latency, meaning only a fraction of this interval should be used for transmission, inflating the throughput requirements. Specifically, a \SI{1}{\Gbps}, \SI{100}{\Hz} video stream requires a \SI{10}{\Gbps} connection to be delivered at a \SI{1}{\milli\second} per-image latency~\cite{struye2020towards}. Non-uniform image sizes due to encoding further inflate this requirement. Finally, per-image reliability must also be high. Packet loss can cause image degradation or even loss of an entire image, severely reducing the \gls{QoE}. This traffic pattern is known as \gls{HRLLC}~\cite{hrllc}. To provide such performance, consistently high-gain links are essential.

A final front involves the \gls{HMD}'s sensors. A modern \gls{HMD} can sense its own position and orientation accurately in real-time. Modern \glspl{HMD}, such as the Meta Quest 2, can generate these measurements without additional equipment, through \textit{inside-out} tracking. This information can be leveraged for fast, accurate beamforming. General beamforming solutions commonly rely on some sort of search algorithm to determine a high-gain beam, possibly from a predefined set of beams~\cite{compressive,OScan,VirtualHierarchical}. As the steering vector is easily derived from pose information, a beam focused in the \gls{LoS} direction can be synthesized directly and efficiently by simply setting the \gls{AWV} in \eqref{eq:awv} to the steering vector in \eqref{eq:steervec}. This maximizes the gain in \eqref{eq:losgain}, as $\left|\mathbf{b}^\mathrm{H} \mathbf{a}\right| \leq 1$ and $\mathbf{a}^\mathrm{H} \mathbf{a} = 1$ for any normalized $\mathbf{a}$ and $\mathbf{b}$.

These specifics of interactive wireless \gls{VR} over \gls{mmWave} indicate the challenges and opportunities for designing an application-specific beamforming solution. As receive-side beamforming, especially during fast rotations, is significantly more challenging than transmit-side beamforming, this work focuses on this first type. While first-generation IEEE 802.11ad hardware often foregoes receive-side beamforming, instead employing quasi-omnidirectional receive beams~\cite{11adCots}, this is inefficient from a link \gls{SNR} perspective~\cite{neighbordisc}, as receive-side beamforming reduces interference~\cite{interferenceMitigation}. This is especially important in multi-\gls{AP} scenarios, where different \glspl{AP} may serve different nearby users. As such, transmit-only beamforming is not expected to achieve the \gls{SNR} needed to fulfill the \gls{HRLLC} requirements. Furthermore, theoretical quasi-omni beams are unrealistic, with gain across their supposedly flat range often varying significantly~\cite{MIDC,BoonAndBane}.

\subsection{Antenna Array Design}\label{sec:arraydesign}
The antenna array for the \gls{HMD} should be designed with the expected environment in mind. This subsection provides general design guidelines.

First of all, we eliminate hybrid and digital arrays. To enable \textit{non-analog} beamforming on a battery-powered device, lower-resolution analog-to-digital and digital-to-analog converters are necessary~\cite{digitalBF}. This reduces achievable \gls{SNR}, making it undesirable for this application. While non-analog arrays are a requirement to send different data streams in different directions, sending/receiving the \textit{same} data stream in/from different directions is perfectly possible with an analog array, using virtual sub-arrays. The potential benefit of sending multiple data streams to the same \gls{HMD} using non-analog arrays is known to be limited~\cite{hbfFullStack}. As such, we opt to rely on analog \gls{HMD}-side arrays.

A next trade-off to consider is between the number of elements in the array, and the spacing between these elements. For a square $N$-element \gls{URA}, the attainable beamwidth in radians, for both azimuth and elevation, is approximately
\begin{equation}
    b_\alpha = \frac{0.886 \lambda}{\sqrt{N}d\cos\alpha}\label{eq:bw}
\end{equation}
with $\alpha$ the steering angle for either azimuth ($\theta$) or elevation ($\psi$), $\alpha=0$ being broadside, and $d$ the inter-element spacing~\cite{phaseBook}. A square \gls{URA} is evident, as the $x$ and $y$-directions of the array, with broadside towards the ceiling, are essentially interchangeable for a user able to rotate around their own axis. The beamwidth equation implies that, for a fixed physical area, adding more elements within said area will not tighten the beamwidth. As such, an inter-element spacing of $d=0.5\lambda$ is often used, as a tighter spacing leads to impractically large beams, while wider spacing is known to create \textit{grating lobes}; undesired side lobes with a directional gain as high as the main lobe's~\cite{phaseBook}. This rule of thumb, however, no longer applies when using interleaved sub-arrays. If only each $a^{th}$ element belongs to the same sub-array, the inter-element spacing within a sub-array effectively becomes $ad$, as illustrated by Fig.~\ref{fig:interleaved}. As such, the physical inter-element spacing should be chosen with a specific $a$ in mind. Furthermore, for sufficiently large arrays, the multi-block layout, as previously illustrated in Fig.~\ref{fig:hybrid}, becomes feasible.

\section{CoVRage} \label{sec:algo}
In this section, we provide a step-by-step explanation of how coVRage works. We first cover how to determine the directions of the sub-beams, how to synthesize them and how to avoid destructive interference along the predicted trajectory between them. Next, we present how to map the sub-beams to the available sub-arrays, which is of crucial importance for multi-block arrays. Finally, we outline the computational complexity of the approach.

\subsection{Sub-beam Generation}
CoVRage must convert measured current and predicted future \gls{HMD} orientations to a set of phase shifts for the phased array in the \gls{HMD}. We decompose this process into three distinct steps. First, we determine how the \gls{AP} appears to move relative to the \gls{HMD}, the reference point. Specifically, we determine the direction of the \gls{AP} at the start and end of the rotation between \gls{HMD} orientations, and the shortest trajectory between these directions. Next, we determine a set of beams that covers this trajectory, achievable by the phased array. Finally, we minimize the destructive interference between the sub-arrays on the trajectory, to avoid having ``blind'' spots along the trajectory. In this section, we represent rotations (and orientations) as \textit{unit quaternions} (e.g., $\mathbf{q}$), as they are easily composed and interpolated. 

\subsubsection{Trajectory Generation}\label{sec:beamtraj}
In the physical environment, the \gls{AP} is a static object, while the \gls{HMD} and its antenna array rotate. From the antenna array's point of view however, the \gls{AP} then appears to rotate around it. In other words, we consider the horizontal coordinate system (i.e., elevation-azimuth system) with its origin in the $(0,0)^\mathrm{th}$ array element, as was used for beamforming in Section~\ref{sec:sysmod}. Every real-world head movement then rotates this coordinate system. In the field of 3D kinematics, this is called a \textit{passive} (or \textit{alias}) transformation~\cite{rotationBook}. For coVRage, an \textit{active} (or \textit{alibi}) transformation, in which we assume the \gls{AP} moves around the \gls{HMD}, is more convenient. Negating the rotational angle converts between active and passive transformations. In other words, when the \gls{HMD} rotates from position $\mathbf{q}$ to $\mathbf{p}$, the coordinate system performs a passive rotation $\mathbf{p}\mathbf{q}\compconj$, equivalent to the \gls{AP} (and any other static objects in the room) performing the active rotation $(\mathbf{p}\mathbf{q}\compconj)\compconj = \mathbf{q}\mathbf{p}\compconj$. To translate rotations to the \gls{AP}'s absolute direction, the \gls{AP} direction at one instant must be known. This can be pre-configured, or measured using existing \gls{AP} sensing approaches~\cite{Pia}.

As the \gls{HMD} is only expected to measure and predict the start and end of the expected rotation within some brief time-frame, coVRage is responsible for generating the path of the \gls{AP} direction during the rotation, between those two points. Given a human head's moment of inertia~\cite{inertia}, we will assume that the most likely trajectory between two orientations is the one with the shortest angular distance. This single rotation, known to exist from Euler's rotation theorem, is easily calculated when the end points are represented as quaternions, using the Slerp algorithm~\cite{Slerp}. When rotating from $\mathbf{q}$ to $\mathbf{p}$, the set of quaternions $(\mathbf{p}\mathbf{q}\compconj)^a$ for $a \in [0,1]$ covers exactly all orientations along the shortest trajectory between the start and end orientations.

\subsubsection{Sub-Beamforming}\label{sec:algo2}
\begin{figure}[!t]
    \centering
    \begin{minipage}{1.0\linewidth}
    \subfloat[Euler Angles]{\includegraphics[width=0.482\textwidth]{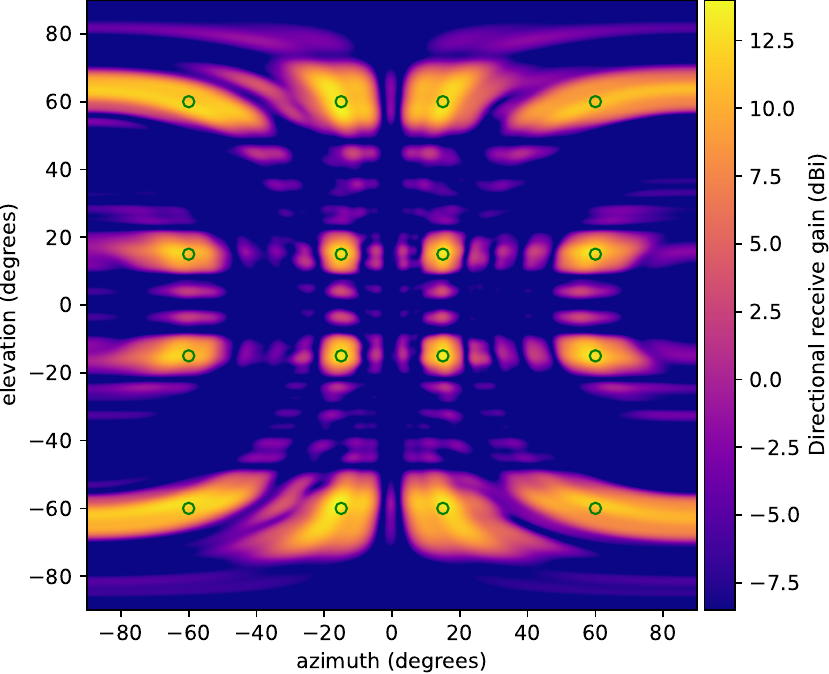} \label{fig:spread1}}
    \subfloat[UV-coordinates]{\includegraphics[width=0.482\textwidth]{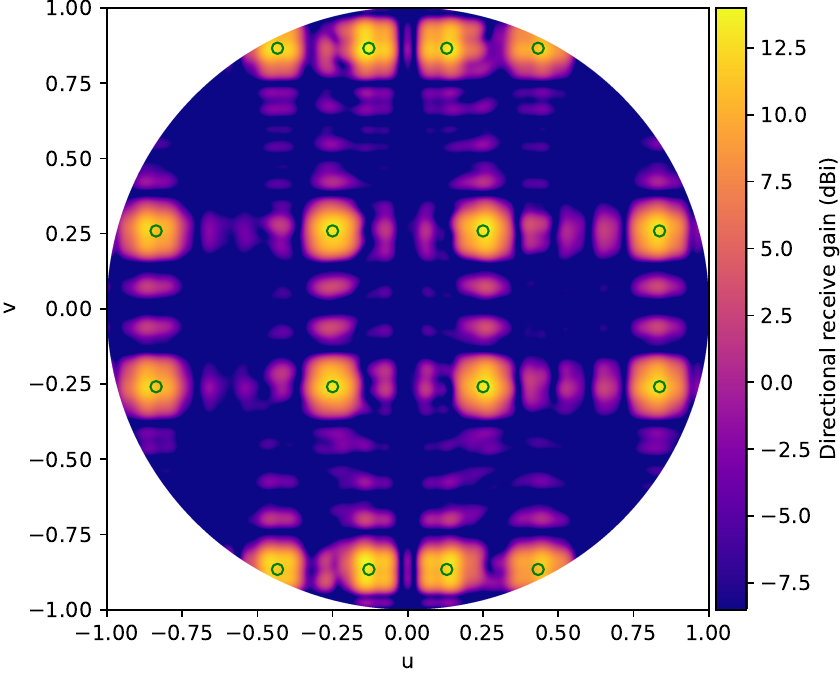} \label{fig:spread2}}
    
    \caption{16 beams, with $\theta$ and $\psi \in \{-60,-15,15,60\}$ degrees, all appear as near-perfect circles in UV-space, but are warped significantly near the edges in Euler coordinates. Green circles show the intended directions.}
    \label{fig:spread}
    \end{minipage}%
\end{figure}

Once the \gls{AP} trajectory as seen from the \gls{HMD} is determined, the algorithm needs to synthesize a beam covering it. For this step, we represent directions not as the commonly used azimuth-elevation pair $(\theta,\psi)$, but rather in UV-coordinates $(u,v)$ with $u = \cos{\psi}\sin{\theta}$ and $v = \sin{\psi}$~\cite{OScan}. While beamwidth depends on the angular distance from broadside for azimuth-elevation (see \eqref{eq:bw}), the beamwidth is nearly invariant to the beam's direction with UV-coordinates~\cite{UVcoords}. Specifically, the beamwidth in UV-space can be approximated by the constant
\begin{equation}
    b_{uv} = \frac{0.886 \lambda}{Nd}\label{eq:bw_uv}
\end{equation}
with an error always under \SI{2}{\percent}, highest near the edges of the hemisphere. As shown in Fig.~\ref{fig:spread}, a rectangular sub-array's beam anywhere in UV-space is as such accurately represented by a circle of constant radius, eliminating the need for complicated, time-consuming beam shape calculations. As conversion from quaternions, used in the previous step, to UV-coordinates, is not a linear transformation, equidistantly sampled quaternion orientations are not necessarily equidistant in UV-space, and are not necessarily on a straight line. Still, the problem of trajectory coverage with sub-beams is essentially reduced to covering a curve using circles. We can sample points along this curve, but the total length of the curve is not (accurately) known a priori. As such, we propose the following sampling approach. We begin by sampling orientations with $a \coloneqq 0$, increasing $a$ by some $\delta_{a}$ for every next sample. In addition, we set some static target UV-space distance between samples $\overline{\delta}_{s}$. For every new sample, we update
\begin{equation}
    \delta_{a} \coloneqq \delta_{a} \frac{\overline{\delta}_s}{\delta_s}
\end{equation}
with $\delta_s$ the actual distance between the two most recent samples. For a sufficiently small $\overline{\delta}_{s}$, this approach generates a near-equidistant list of directional samples along the trajectory. Using these samples, we now easily estimate the total trajectory length $t$ and the desired angular distance between two adjacent sub-beams $\Delta_s$ as
\begin{equation}
    \Delta_s = \frac{t}{s-1}
\end{equation}

where $s$ is the number of sub-beams available. By using $s-1$ rather than $s$, the centers of the first and last sub-beam, rather than their edges, can coincide with the start and end of the trajectory. Experimentation showed that this was necessary to achieve sufficient gain near the start and end of the trajectory. It is then straightforward to aim the sub-beams by iterating over all sampled trajectory points and aiming a sub-beam towards one whenever the desired distance $\Delta_{s}$ is reached. The points halfway between each adjacent pair of sub-beams are called \textit{midpoints} and are stored for the following step. Note that an array can only cover a trajectory if $\Delta_{s} \leq b_{uv}$.

The optimal \gls{AWV} $\mathbf{w}$ pointing to $(\theta,\psi)$, ignoring \gls{NLoS} paths, is easily derived from the channel model. The gain in some \gls{AoA} $(\theta,\psi)$ is maximized by setting the \gls{AWV} (before normalization) equal to the steering vector for that \gls{AoA}, defined in \eqref{eq:steervec}. For the full \gls{AWV} of coVRage's entire beam, which aims at several directions, it suffices to, for each element $(x,y)$, set its \gls{AWV} element $w_{x,y}$ to the steering vector for the \gls{AoA} of the sub-beam formed by the sub-array that element belongs to. In other words, the beamforming parameters of $w_{x,y}$ shown in \eqref{eq:awv_el} are set to $\alpha_{x,y} \coloneqq 1$ and
\begin{equation}
\phi_{x,y} \coloneqq \frac{2\pi d}{\lambda}\left(x\sin{\theta_s}\cos{\psi_s}+y\sin{\psi_s}\right)
\end{equation}
when the element at position $(x,y)$ contributes to the sub-beam aimed at $(\theta_s,\psi_s)$.
\begin{figure}[!t]
    \centering

    \begin{minipage}{1.0\linewidth}
    \subfloat[Same]{\includegraphics[width=0.231\linewidth]{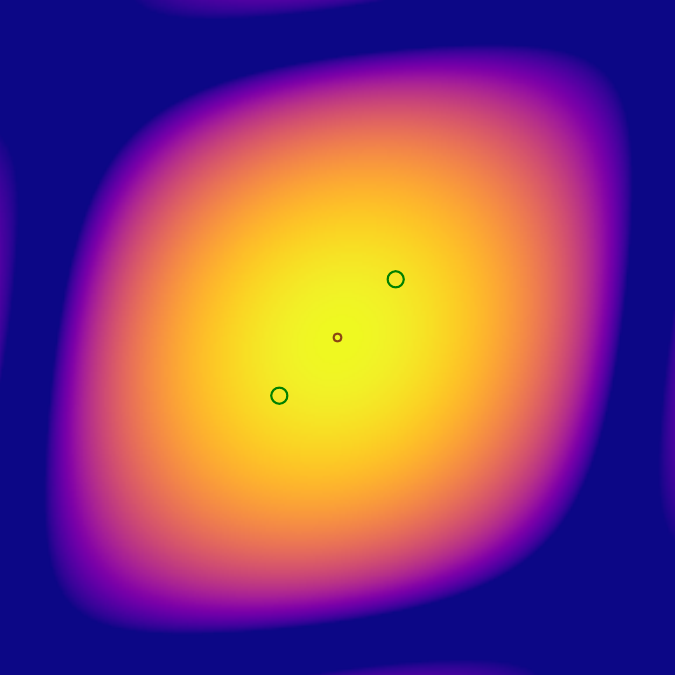} \label{fig:blockdiff_same}}
    \subfloat[Hor.]{\includegraphics[width=0.231\linewidth]{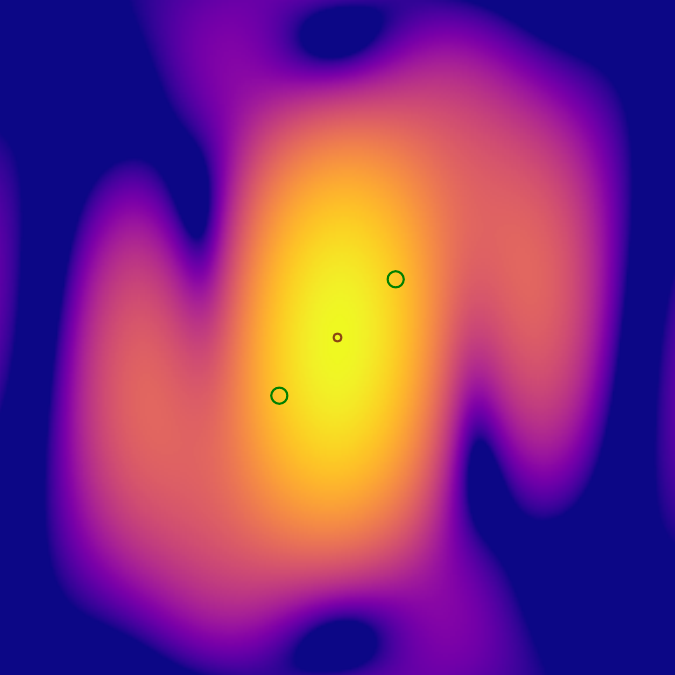} \label{fig:blockdiff_hor}}
    \subfloat[Vert.]{\includegraphics[width=0.231\linewidth]{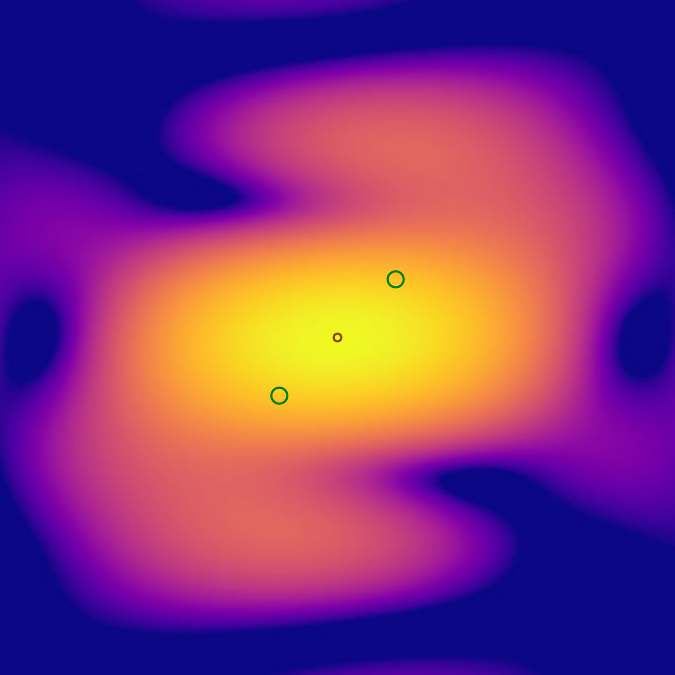} \label{fig:blockdiff_vert}}
    \subfloat[Diag.]{\includegraphics[width=0.231\linewidth]{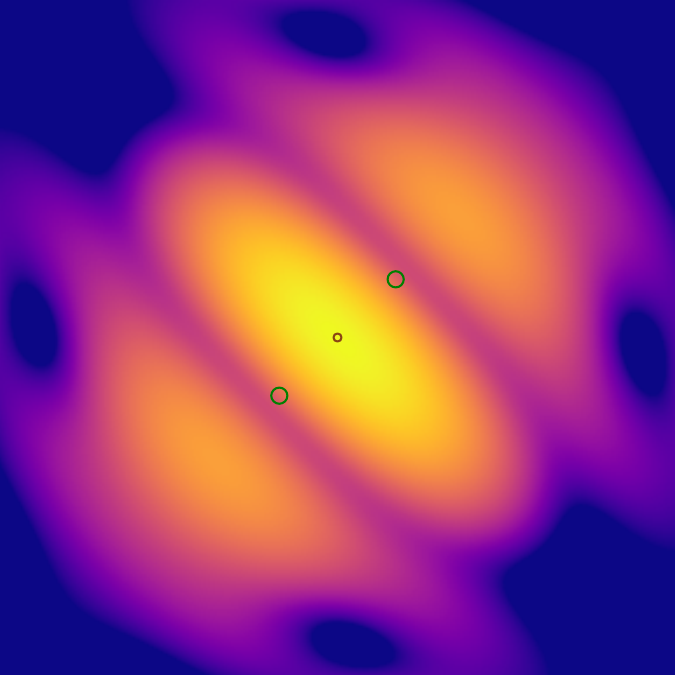} \label{fig:blockdiff_diag}}
    \caption{Aperture fusion is clearly visible when forming the same two adjacent sub-beams (large green circle) with the same midpoint (small brown circle) with sub-arrays either in the same block, or in horizontally/vertically/diagonally different blocks.}
    \label{fig:blockdiff}
    \end{minipage}
\end{figure}
\subsubsection{Sub-Beam Syncing}
Once the sub-array layout is determined and each sub-array is aimed properly, the remaining step is to synchronize the sub-beams, eliminating destructive interference between sub-arrays along the trajectory. As global optimization at this level is challenging and expensive, we apply a heuristic inspired by previous work on analog sub-arrays~\cite{FlexibleCoverage}. Specifically, we minimize destructive interference between adjacent sub-beams where it is expected to be the most impactful. In this case, this is the point along the trajectory equidistant from the two sub-beam directions. These midpoints were already stored during the previous step. To maximize constructive interference between adjacent sub-beams at the midpoint, coVRage calculates the phase difference between the two, and applies this as an additional phase shift to all elements contributing to the second sub-beam, making the two sub-beams phase-aligned at the midpoint. The phase of a sub-beam $i$ for midpoint $m = (\theta_m,\psi_m)$ is
\begin{equation}
\phi_{i,m} = \angle \left(\mathbf{w}_i^\mathrm{H}\mathbf{a}(N_R, \theta_m, \psi_m)_i\right)
\end{equation}
 where the subscript $i$ indicates that all entries not corresponding to an array element contributing to sub-array $i$ are set to 0.
 For all sub-arrays $i>0$, all entries of $\mathbf{w}$ corresponding to an element contributing to that sub-array are multiplied by a phase offset
 \begin{equation}
    o_i = \exp{\left(j \left(\phi_{i-1,i-1} - \phi_{i,i-1}\right)\right)}
 \end{equation}
where $o_{i+1}$ must only be calculated after $o_i$ was applied.
\subsection{Sub-beam Assignment}
With a fully interleaved array, the mapping of sub-beams to the available sub-arrays does not influence performance. With a multi-block array however, this becomes an important aspect of the algorithm. In a multi-block array, two sub-arrays from the same block interact as if they are in an interleaved configuration, while sub-arrays from different blocks interact as if localized. Fig.~\ref{fig:hybrid} shows a multi-block array configured to consist of 4 blocks of $4 \times 4$ elements each.

As \eqref{eq:bw} implies, beamwidth is inversely proportional to the array aperture, i.e., the physical size of the array. When sub-arrays behave as interleaved, combining two by aiming them in the same direction increases the total aperture by only one inter-element spacing $d$, regardless of array size. In the localized case however, the total aperture (roughly) doubles, effectively halving the beamwidth. When two sub-beams are only slightly offset, this \textit{aperture fusion} still occurs to a large extent, as Fig.~\ref{fig:blockdiff} illustrates. CoVRage for multi-block arrays must therefore take care to avoid aperture fusion where it could cause ``holes" in the coverage, but could also leverage this effect for tighter coverage if desired. We therefore propose two different approaches to multi-block beamforming: a \textit{loose} beam where aperture fusion is avoided as much as possible as to maintain beamwidth offered by a single-block array, and a \textit{tight} beam where aperture fusion is triggered as much as possible as to achieve higher gain exactly on-trajectory. In both cases, further measures must be taken to ensure a stable coverage of consistent width throughout the trajectory. In what follows, we assume that the considered trajectory is predominantly \textit{horizontal}, meaning its range on the $u$-axis exceeds that on the $v$-axis. For the vertical version, simply swap the $x$ and $y$-coordinates of the array elements.
\begin{figure*}[!t]
    \centering
    \begin{minipage}{1.0\linewidth}
    \centering
    \subfloat[Default case, 2 distinct blocks]{\includegraphics[width=0.485\linewidth]{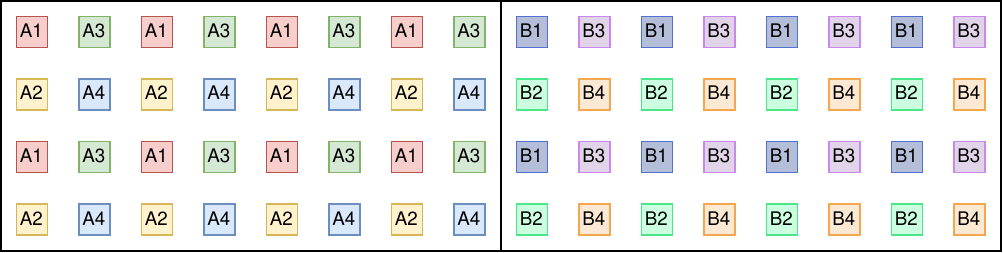} \label{fig:halfwayno}}
    \hfill
    \subfloat[4 sub-arrays reduced to 2 transitional sub-arrays]{\includegraphics[width=0.485\linewidth]{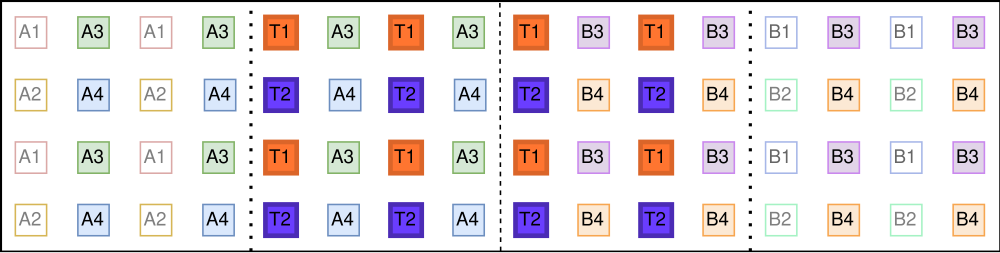} \label{fig:halfwayyes}}
    \end{minipage}
    \caption{On the left, a $16 \times 4$ antenna array is configured as two $8 \times 4$ blocks of four $4 \times 2$ interleaved sub-arrays each. To reduce aperture fusion between sub-arrays from different blocks, we introduce transitional sub-arrays in the right configuration, shown with thicker edges. Half of the elements initially assigned to sub-arrays A1 and B1 now form a new transitional sub-array T1, in a new transitional block (delimited by dotted lines), overlapping with the two original blocks (separated by dashed line). The remaining elements of A1 and B1 are disabled, shown in white. Analogously, T2 is formed from A2 and B2.}
    \label{fig:halfway}
\end{figure*}

We first present a method of mitigating unwanted aperture fusion. For this, we introduce the \textit{transitional sub-array}, illustrated in Fig.~\ref{fig:halfway}. To form a horizontal transitional sub-array, take two sub-arrays from horizontally adjacent blocks, and combine them into a new sub-array of the same size by disabling the leftmost half of the left sub-array, and the rightmost half of the right sub-array. Disabling individual antenna elements requires 1-bit amplitude control, which is often assumed to be supported~\cite{deactivate,VirtualHierarchical}. Using a transitional sub-array, aperture fusion is smoothed out over a larger angular distance. This is easily derived from the beamwidth equation in \eqref{eq:bw_uv}. Consider two sub-arrays, forming two adjacent sub-beams. These will be phase-aligned at the midpoint, meaning \eqref{eq:bw_uv} can be used to calculate beamwidth as if the involved antenna elements formed one larger sub-array aimed towards this midpoint. As such, having two sub-arrays in different blocks, instead of the same block, will reduce midpoint beamwidth by \SI{50}{\percent}. When instead using a transitional sub-array, this reduction is limited to \SI{33}{\percent}.

As sub-arrays within the same block are interchangeable, the main challenge concerns \textit{block transitions}. Each time two adjacent sub-beams are mapped to sub-arrays in different blocks, either horizontally or vertically, a block transition occurs, triggering aperture fusion. To optimize performance, we must decide whether to minimize or maximize the number of transitions for the two directions. If aperture fusion is desirable for one direction, a transition can occur for every pair of adjacent sub-beams, and only a single transition is needed in the undesirable direction, for a $2 \times 2$ block array. If both directions are to be minimized, a minimum of three transitions must occur, including at least one in each direction. For any undesirable transition, we must also determine an appropriate number of transitional sub-arrays. As each transitional sub-array deactivates elements and decreases the total sub-array count, this number should be minimized.
\begin{figure*}[!t]
    \centering
    \begin{minipage}{1.0\linewidth}
    \subfloat[Tight]{\includegraphics[width=0.2365\textwidth]{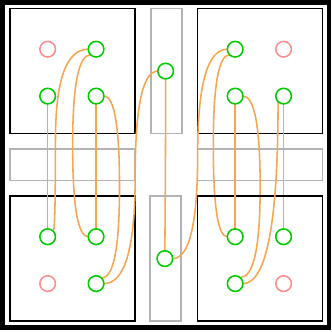} \label{fig:layouttight}}
    \hfill
    \subfloat[Loose, non-diagonal]{\includegraphics[width=0.2365\textwidth]{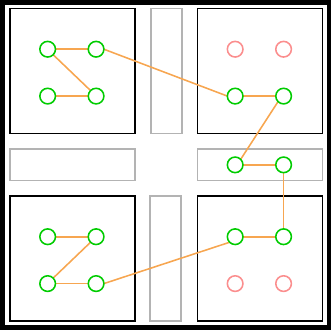} \label{fig:layoutloosenodiag}}
    \hfill
    \subfloat[Loose, semi-diagonal]{\includegraphics[width=0.2365\textwidth]{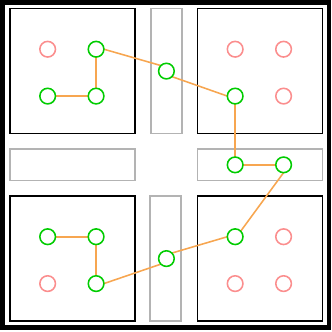} \label{fig:layoutlooseinterm}}
    \hfill
    \subfloat[Loose, diagonal]{\includegraphics[width=0.2365\textwidth]{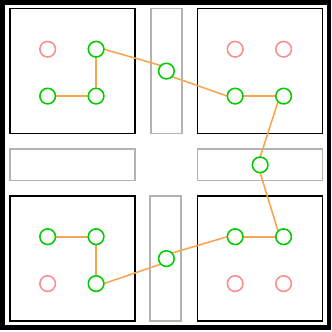} \label{fig:layoutloosediag}}
    \end{minipage}
    \caption{The order in which sub-arrays are mapped to sub-beams, for each type of beam, for an array such as seen in Fig.~\ref{fig:hybrid}. Each circle is a sub-array, consisting of several antenna elements. Each sub-array may be active (green) or inactive (pink). The orange line connects all active sub-arrays in order, such that connected sub-arrays form adjacent sub-beams. For each transitional sub-array (in a grey box), a sub-array is deactivated in each adjacent block. Within a block, sub-arrays are interchangeable.}
    \label{fig:beamlayouts}
\end{figure*}
\begin{table}[t]
    \caption{Loose (horizontal) beam configurations}
    \begin{center}
    \begin{tabular}{cccc}
    \toprule
    &\multicolumn{2}{c}{\textbf{Trans. sub-arrays}} & \\
    \cmidrule(lr){2-3}
    & \textbf{Hor.} & \textbf{Vert.} & \textbf{Sub-beams} \\
    \cmidrule(lr){2-4}
    \textbf{Non-diag.} & 0 & 2 & 14 \\
    \textbf{Semi-diag.} & 1 & 2 & 12 \\
    \textbf{Diag.} & 1 & 1 & 13 \\
    \bottomrule
    \end{tabular}
    \label{tab:loose}
    \end{center}
\end{table}

For a (horizontal) tight beam, vertical transitions tighten the beam vertically and are therefore desirable. Horizontal transitions may cause coverage holes and are therefore minimized. Extensive experimentation showed that two transitional sub-arrays were needed to consistently eliminate said holes. This leaves a total of 14 sub-beams. Fig.~\ref{fig:layouttight} illustrates this layout.

With a loose beam solution, all block transitions are undesirable. With horizontal trajectories, we opt for two horizontal transitions and one vertical transition, as the latter has a larger impact on coverage consistency. Careful experimentation revealed that the number of required transitional sub-arrays depends on the slope of the trajectory. We broadly identified three different cases depending on (mean) UV-space slope $m$ of the trajectory: a \textit{non-diagonal} trajectory with $\vert m\vert < 1/3$, a \textit{diagonal} trajectory with $\vert m\vert > 2/3$ and a \textit{semi-diagonal} trajectory in-between. Table \ref{tab:loose} shows the number of transitional sub-arrays needed per transition, and the resulting number of sub-beams, and Fig.~\ref{fig:layoutloosenodiag} to \ref{fig:layoutloosediag} illustrate them. As the number of transitional sub-arrays is highest in the semi-diagonal case, the other two can be seen as optimizations.

\subsection{Computational Complexity}
Given that analog beamforming with quantized phase shifters is generally NP-hard~\cite{nphard}, we opted for a heuristic approach for coVRage, intended to run in real-time. In this subsection, we analyze the complexity of the algorithm's main steps. The complexity of trajectory generation depends on the trajectory length $t$ and the parameter $\overline{\delta}_{s}$, controlling the target distance between sampled points, resulting in a complexity of $O(\frac{t}{\overline{\delta}_{s}})$. If needed, one can increase the parameter $\overline{\delta}_{s}$ to reduce runtime. The only downside of this is an increase of $\frac{\overline{\delta}_{s}}{2}$, the worst-case angular difference between the selected and the optimal sub-beam direction. Next, sub-beam synthesis is $O(N)$ with $N$ the number of elements, which is optimal when not relying on pre-computed beams from a codebook. Finally, sub-beam syncing is $O(s)$ with $s$ the number of sub-beams. Based on the runtime of the current, unoptimized implementation, we fully expect the algorithm to synthesize a beam in under \SI{1}{\milli\second} on-\gls{HMD}. As the \gls{HMD}'s processor should have some idle time between processing arriving content frames (once every \SI{8.33}{\milli\second} at \SI{120}{\hertz}), we do not expect coVRage's computations to have any impact on \gls{QoE}. Furthermore, as we target a prediction horizon in the order of \SI{100}{\milli\second}, the runtime latency is not expected to impact coVRage's effectivity.
\section{Evaluation}\label{sec:eval}
In this section, we simulate coVRage to evaluate how well it performs in the envisioned scenario. First, we visualize its performance in individual trajectories to illustrate that it achieves its trajectory-covering goal with both loose and tight beams. Then, we evaluate its performance over many thousands of randomly generated trajectories using carefully crafted metrics to show the algorithm's consistency. In the evaluation, we consider antenna arrays that would fit into an \gls{HMD}. Measuring several commercially available devices showed that a \SI{4}{\centi\metre} $\times$ \SI{4}{\centi\metre} array is a fair assumption. For the first set of experiments, we simulate a \SI{120}{\giga\hertz} array, resulting in a $64 \times 64$ quarter-wavelength-spaced array, divided into four blocks of four interleaved half-wavelength-spaced sub-arrays each, thereby capable of producing 16 distinct sub-beams. For these experiments, we use the simplified \gls{LoS}-only model as its complexity vastly increases the number of simulations we can perform, and assume non-quantized phase shifters. Afterwards, we consider a $32 \times 32$ element array with a single block of four interleaved sub-arrays, targeting \SI{60}{\giga\hertz}, and evaluate the impact of introducing \gls{NLoS} links and quantized phase shifters. All simulations are implemented in C\texttt{++}.
\begin{figure}
    \begin{minipage}{1.0\linewidth}
    \subfloat[Trajectory lengths]{\includegraphics[width=0.485\textwidth]{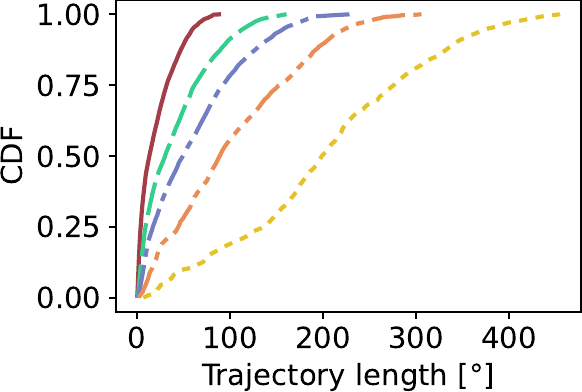}}
    \hfill
    \subfloat[Velocities]{\includegraphics[width=0.485\textwidth]{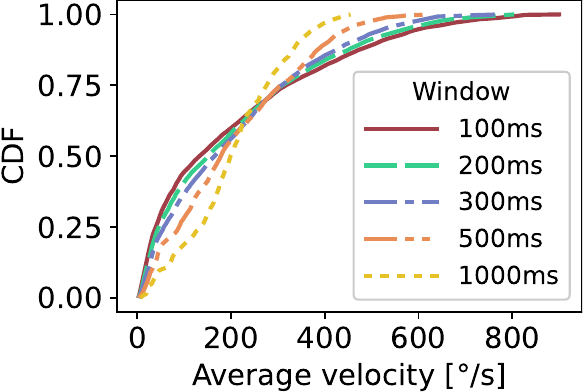}}
    \captionof{figure}{\glsfirst{CDF} of the angular trajectory lengths and velocities within several sliding windows, gathered from real-world measurements.}
    \label{fig:angles}
    \end{minipage}
\end{figure}
\subsection{Individual trajectories}\label{sec:indiv_traj}
Before evaluating coVRage, we investigate the range of trajectories we can reasonably expect to encounter. One existing dataset on high-motion head movement shows a maximal velocity of \SI[per-mode=fraction,fraction-function=\sfrac]{167}{\degree\per\second}~\cite{RotationPrediction}. It is however unclear if this shows instantaneous velocity, or velocity averaged over some temporal window. Another work shows speeds up to \SI[per-mode=fraction,fraction-function=\sfrac]{360}{\degree\per\second} over a \SI{100}{\milli\second} window for only the yaw axis~\cite{OScan}. In addition to this existing data, we gathered a dataset focused on determining maximal realistic angular velocities. A subject wearing the Meta Quest 2 was encouraged to move quickly and erratically as much as they were comfortable with, and rotational data was logged at the maximal frequency of \SI{972}{\hertz}. To mitigate any sample-to-sample inaccuracies in the measurements, we analyze the performed rotations with sliding windows of different lengths, between \SI{100}{\milli\second} and \SI{1}{\second}. Fig. \ref{fig:angles} shows the range of trajectory lengths and velocities produced within sliding windows of different lengths, across \SI{45}{\second} of rapid movement. This shows a maximal rotation of \SI{160}{\degree} for \SI{200}{\ms}. Intuitively, it makes sense that a human is not able to turn their head more than \SI{180}{\degree} in a single rapid motion. The maximum velocity, averaged over the window, also decreases with longer windows, as the human body intuitively cannot sustain maximum angular velocity for long. At \SI{200}{\milli\second}, the velocity peaks at \SI[per-mode=fraction,fraction-function=\sfrac]{804}{\degree\per\second}. We use the \SI{200}{\milli\second} window as a reference, as, for larger sliding windows, fast rotation was usually not sustained within the full window. Based on this, we only consider trajectories of at most \SI{180}{\degree}.
\begin{figure*}
    \centering
        \subfloat[Traj. A (loose)]{\includegraphics[width=0.3\textwidth]{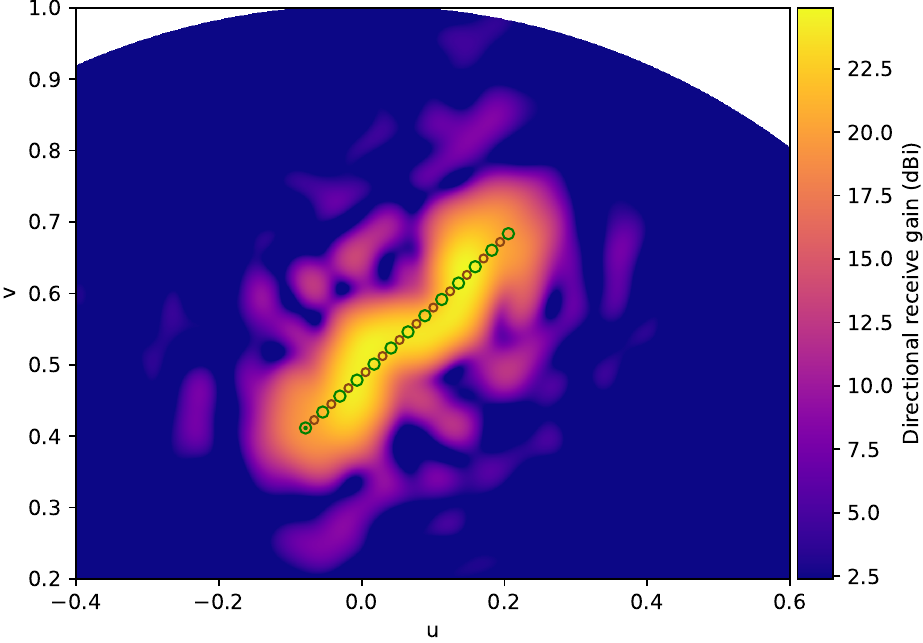} \label{fig:ex1loose}}
        \hfill
        \subfloat[Traj. B (loose)]{\includegraphics[width=0.3\textwidth]{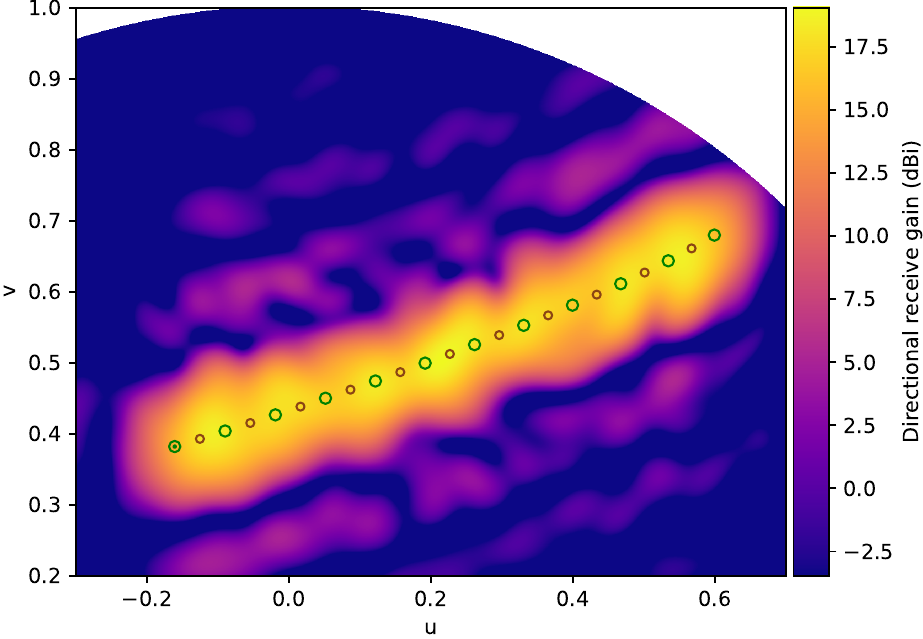} \label{fig:ex2loose}}
        \hfill
        \subfloat[Traj. C (loose)]{\includegraphics[width=0.3\textwidth]{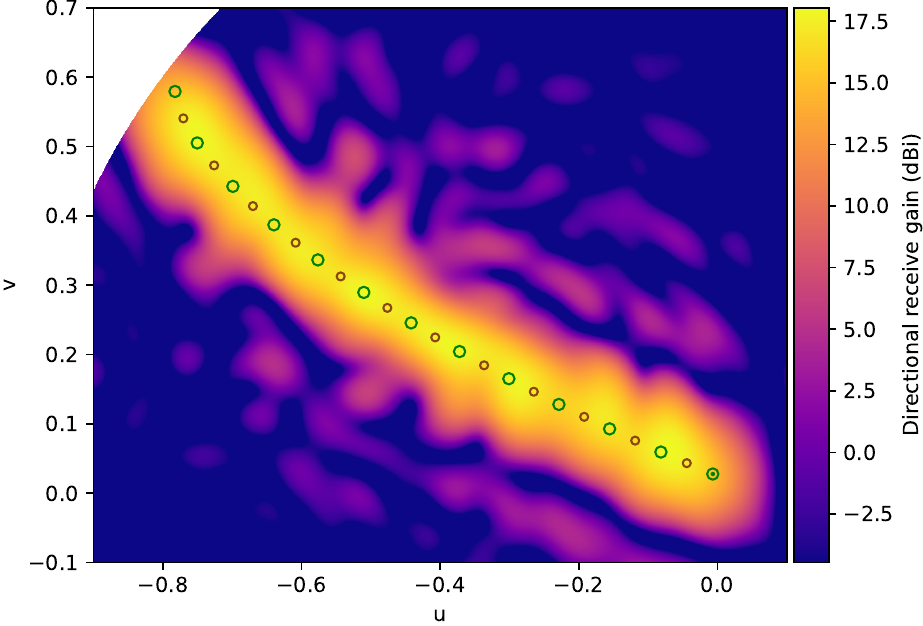} \label{fig:ex3loose}}
        \vspace{-2mm}
        \subfloat[Traj. A (tight)]{\includegraphics[width=0.3\textwidth]{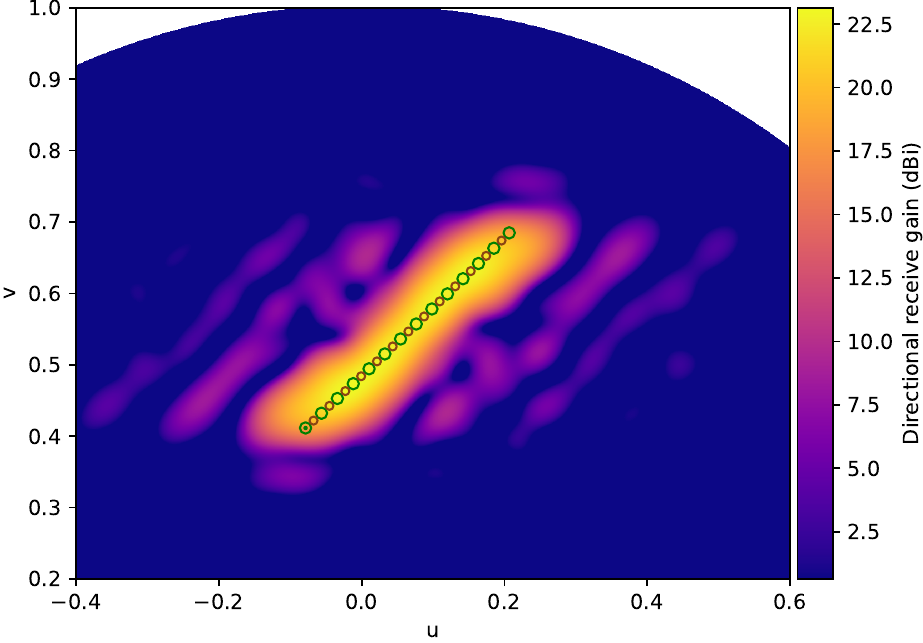} \label{fig:ex1tight}}
        \hfill
        \subfloat[Traj. B (tight)]{\includegraphics[width=0.3\textwidth]{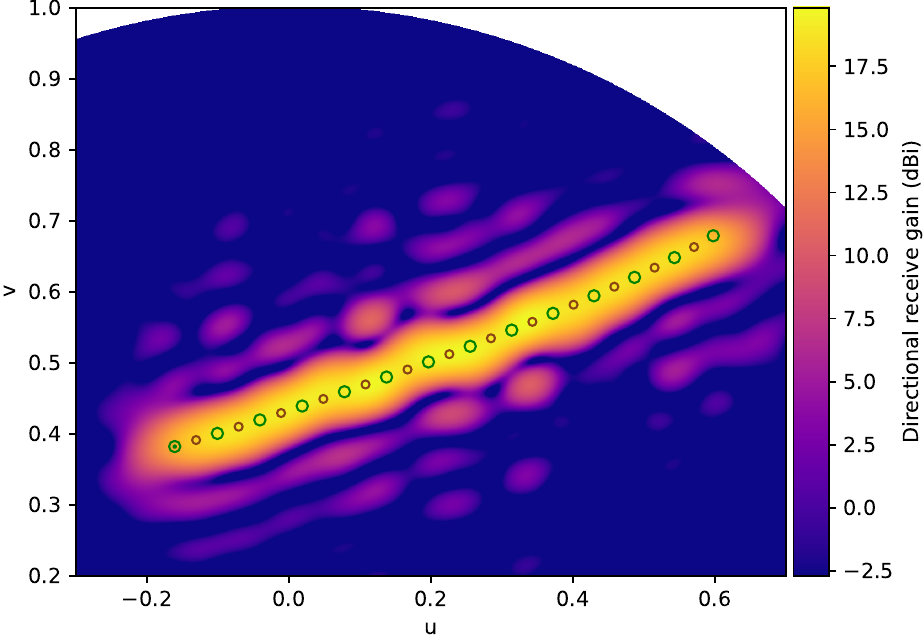} \label{fig:ex2tight}}
        \hfill
        \subfloat[Traj. C (tight)]{\includegraphics[width=0.3\textwidth]{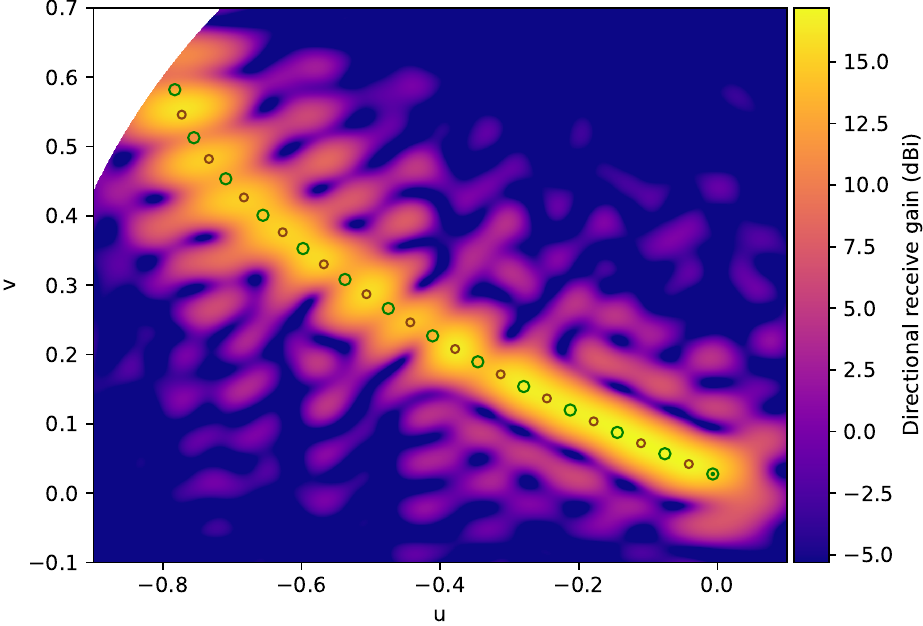} \label{fig:ex3tight}}
        \vspace{-2mm}
        \subfloat[Traj. D (loose)]{\includegraphics[width=0.3\textwidth]{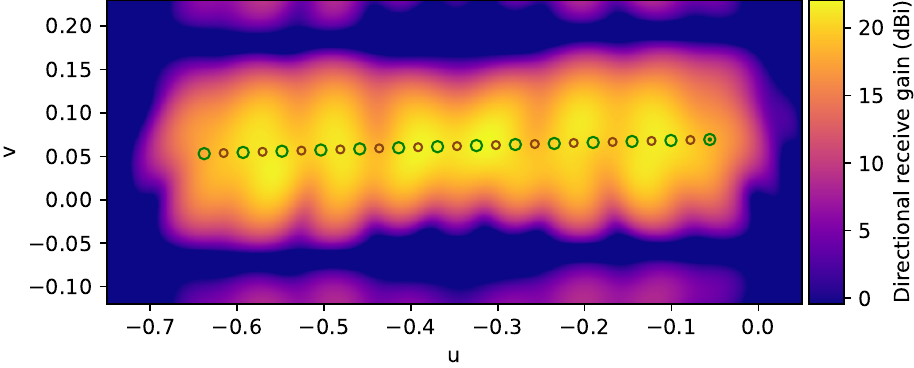} \label{fig:ex4loose}}
        \hfill
        \subfloat[Traj. E (loose)]{\includegraphics[width=0.3\textwidth]{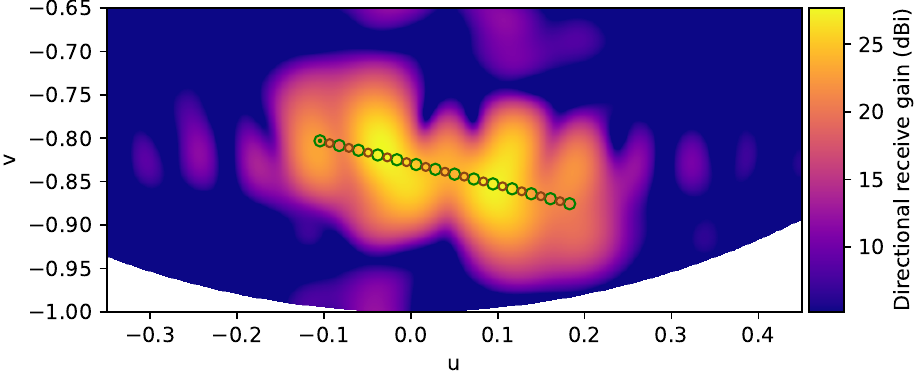} \label{fig:ex5loose}}
        \hfill
        \subfloat[Traj. F (loose)]{\includegraphics[width=0.3\textwidth]{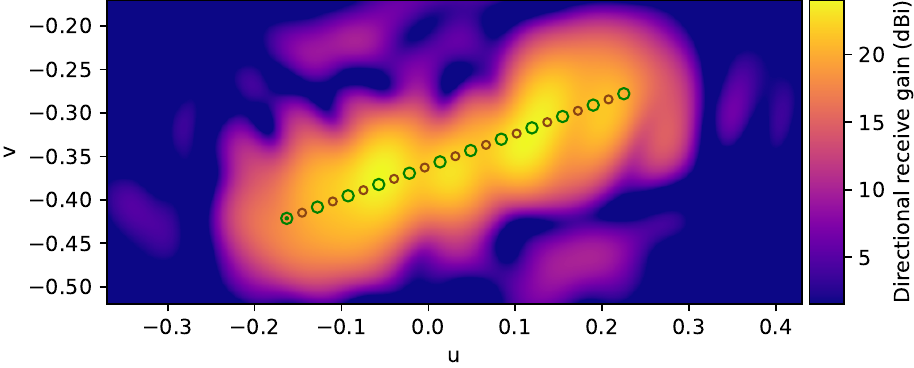} \label{fig:ex6loose}}
        \vspace{-2mm}
        \subfloat[Traj. D (tight)]{\includegraphics[width=0.3\textwidth]{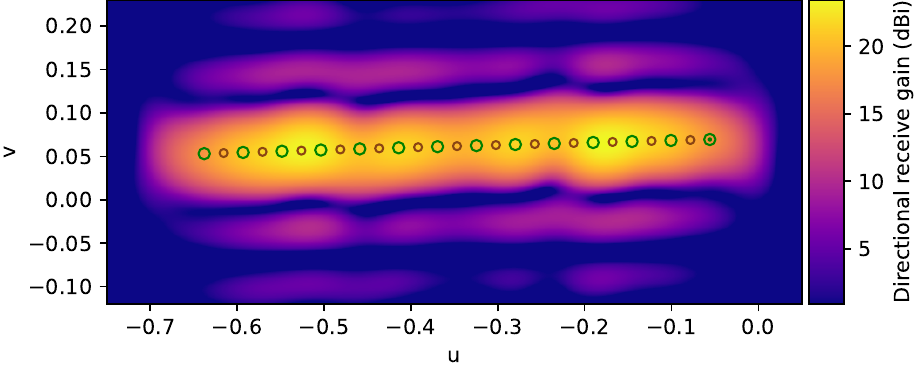} \label{fig:ex4tight}}
        \hfill
        \subfloat[Traj. E (tight)]{\includegraphics[width=0.3\textwidth]{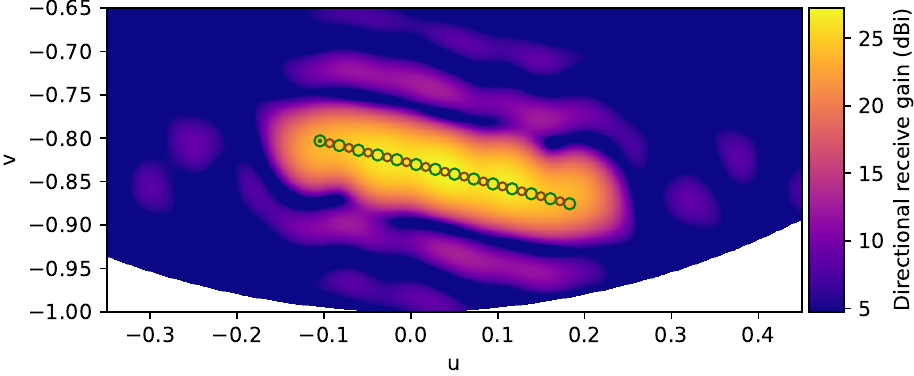} \label{fig:ex5tight}}
        \hfill
        \subfloat[Traj. F (tight)]{\includegraphics[width=0.3\textwidth]{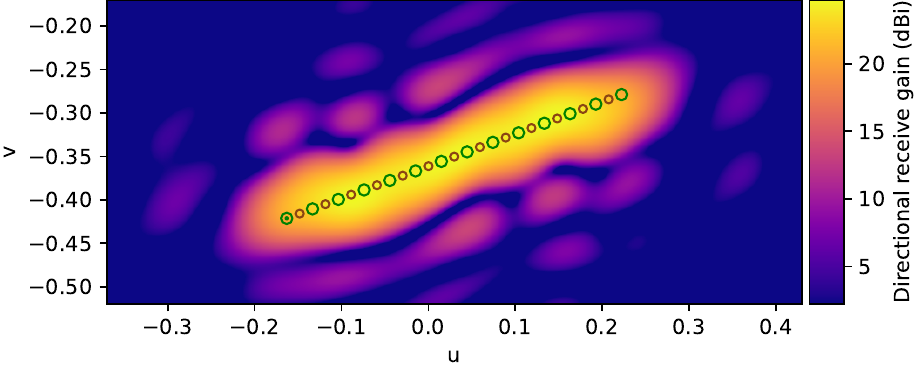} \label{fig:ex6tight}}
        \caption{Loose and tight beam coverage for a broad range of trajectories, showcasing coVRage's effectiveness, as well as known shortcomings (in \ref{fig:ex3tight} and \ref{fig:ex5loose}). Large green circles indicate sub-beam directions, with the double circle being the start of the trajectory. Small brown circles indicate the midpoints, where two adjacent sub-beams are phase-aligned. Plots are zoomed-in versions of the full-range UV plot (see Fig.~\ref{fig:spread2}), with plots in the same row sharing the same zoom level.}
        \label{fig:examples}
\end{figure*}

To visually present the performance of coVRage, we introduce the \textit{gain map}. For some selected \gls{AWV}, the gain map shows, for every possible direction, what the beamforming gain would be if this direction were the \gls{AoA}. The map covers the hemisphere centered around the receive array's broadside, which points directly at the ceiling when the subject is standing up straight. Ideally, this gain should be consistently high along the trajectory and low elsewhere. For an initial analysis, we generate a broad set of gain maps, and pick a subset that shows both the general effectiveness of the algorithm, as well as its main limitations. Fig.~\ref{fig:examples} shows these gain maps for both loose and tight beams, showing that both beam types generally achieve their goals. Two trajectories were selected to highlight known limitations of the algorithm. The tight beam in Fig.~\ref{fig:ex3tight} has reduced gain in the top half. Longer trajectories can curve noticeably in UV space, meaning one may transition from predominantly horizontal to predominantly vertical. As sub-array order varies significantly between the two (Fig.~\ref{fig:layouttight} shows the horizontal version, and can be mirrored diagonally to obtain the vertical one), and only one order can be used throughout the trajectory, the tight beam does not perform well along part of the trajectory. While the tight beam was designed with either horizontal or vertical tightening in mind, it does not degrade significantly for shorter diagonal trajectories such as Fig.~\ref{fig:ex1tight}, although the tightening effect is reduced. As loose beams take the trajectory's slope $m$ into account, the difference in sub-array order for beams with $\vert m\vert$ slightly under 1 (Fig.~\ref{fig:layoutloosediag}) and slightly over 1 (Fig.~\ref{fig:layoutloosediag} mirrored diagonally) is minor: one horizontal transitional sub-array is swapped for a vertical one. As such, loose beams do not struggle in this case, as Fig.~\ref{fig:ex3loose} shows.

Secondly, Fig.~\ref{fig:ex5loose} and, to a lesser extent, Fig.~\ref{fig:ex1loose}, show an inconsistent beamwidth for the loose beam. This occurs in short trajectories, because non-adjacent sub-beams may be close enough to influence each other significantly. In this case, the number of transitional sub-arrays introduced to mitigate aperture fusion may not suffice for very close sub-beams. Overall, the issue with tight beams only occurs for longer trajectories, while the issue for loose beams only occurs with shorter trajectories, with the two susceptible ranges being non-overlapping. As accurate measurements and predictions are more challenging under fast motion, during which erratic changes in velocity or heading are more likely to occur, loose beams are generally preferable for longer trajectories. Conversely, tight beams are generally superior for shorter trajectories, which are easier to measure and predict. As such, each beam type's issue only occurs when that beam type is not the most appropriate beam type, making these issues only a minor limitation of the coVRage approach, which can be solved by intelligently selecting the most appropriate beam type, based on parameters such as expected trajectory length and confidence in predictions.
\subsection{Detailed analysis}
\begin{figure*}[!t]
    \centering
    \begin{minipage}{1.0\linewidth}
    \subfloat[On-trajectory]{\includegraphics[width=0.32\textwidth]{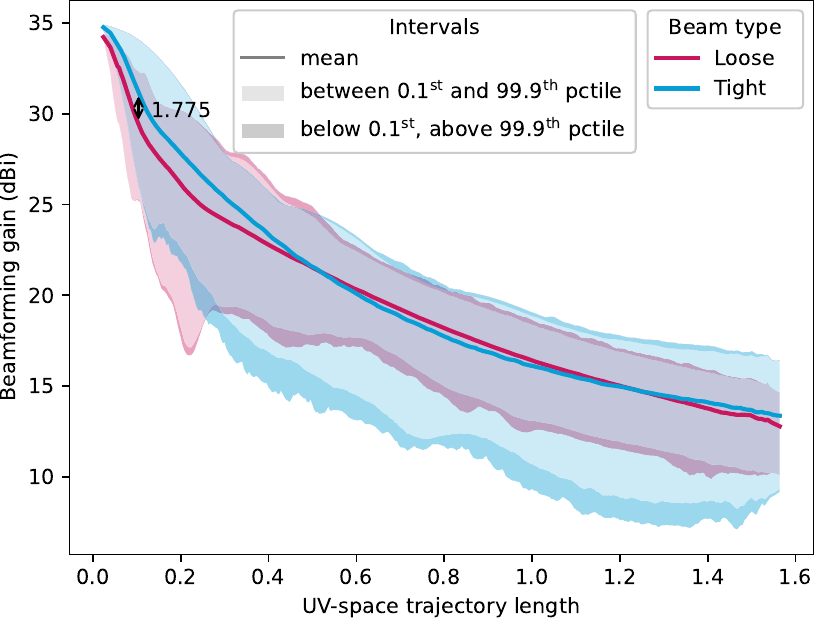} \label{fig:trajgood}}
    \hfill
    \subfloat[Off-trajectory (severe)]{\includegraphics[width=0.326\textwidth]{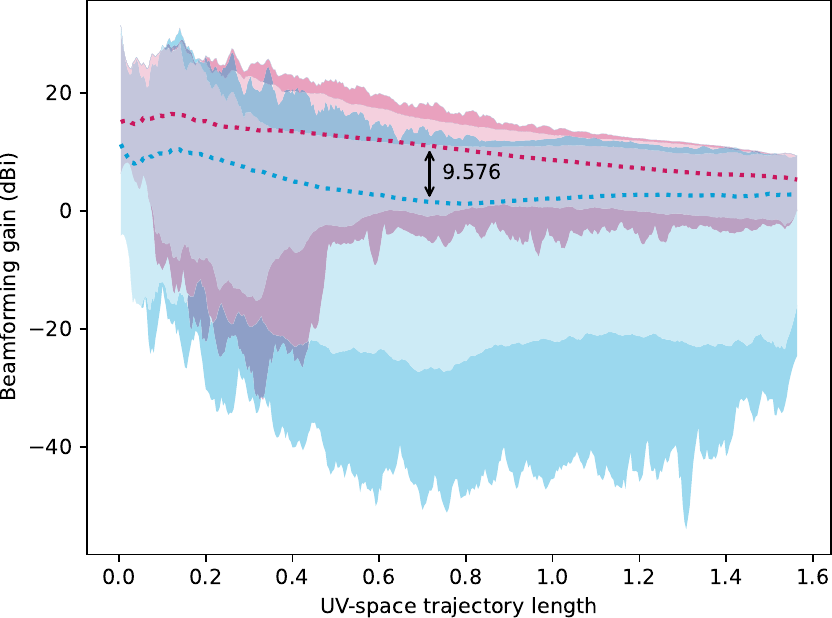} \label{fig:trajbad}}
    \hfill
    \subfloat[On and off-trajectory]{\includegraphics[width=0.3205\textwidth]{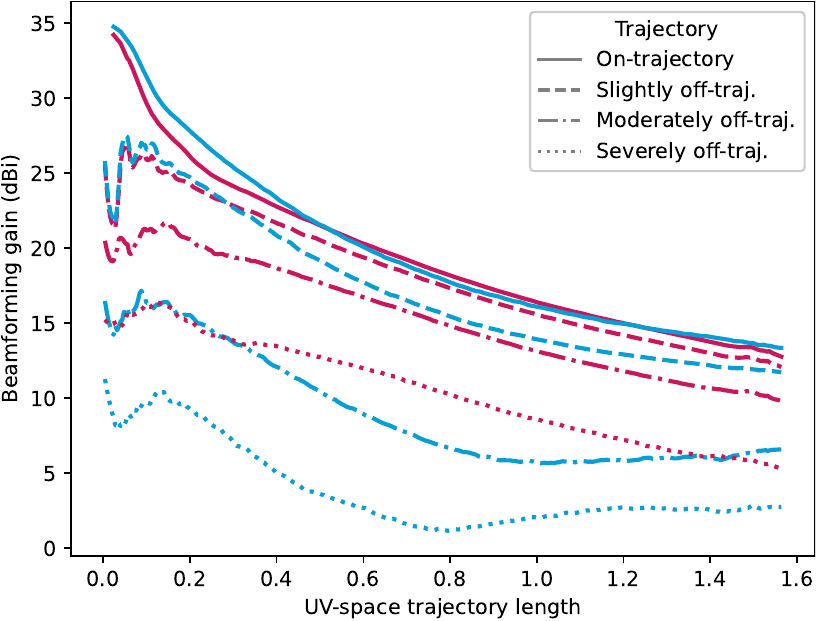} \label{fig:trajall}}
    \end{minipage}
    \caption{Fan charts showing the mean beamforming gain along the full actual trajectory for all beam types depending on trajectory length, aggregated over \num{100000} simulations, grouped by trajectory length. For off-trajectory, the trajectory expected while beamforming differs from the actual trajectory along which gain was measured.}
    \label{fig:trajgoodbad}
\end{figure*}
Next, we evaluate whether coVRage can repeatedly achieve its consistent trajectory coverage for a wide range of trajectories. Even occasional drops in coverage are expected to have a significant negative impact on \gls{QoE}. As such, we randomly generate \SI{100000}{} trajectories, in the \SI{20}{\degree} to \SI{180}{\degree} range. We sample the start and end points of the trajectories uniformly in the 3D rotation group $\mathrm{SO}(3)$ using a well-known algorithm~\cite{randQuatPaper,randQuatBook}. We introduce three metrics which together characterize coVRage's consistency: \textit{trajectory gain}, \textit{gain concentration} and \textit{gain variation}. Trajectory gain measures the gain experienced on all points along the trajectory. As implied by Parseval's theorem~\cite{parseval}, increasing the trajectory length decreases achievable (stable) gain along said trajectory. As such, we report trajectory gain as a function of trajectory length. We subdivide the metric into on-trajectory gain, where the trajectory was perfectly measured and predicted, and off-trajectory gain, where we determine the gain along a trajectory not coinciding with the trajectory coVRage intended to cover, evaluating the impact of imperfect measurements and/or predictions. For the off-trajectory case, every point in the actual trajectory is shifted along the normal in said point by a fixed UV-space distance. Secondly, the gain concentration is based on the linear gain integrated over a range of directions in the hemisphere. Specifically, for some angular distance $\delta$, the gain concentration indicates the fraction of the total integrated gain that is at most $\delta$ removed from the nearest point on the trajectory. This provides an indication of beamwidth. Ideally, this function would increase near-linearly within the intended width of the beam, reaching $1$ at the edge of the intended coverage zone. Any gain outside of the intended coverage could be considered ``lost", following again from Parseval's theorem. Finally, gain variation is the difference between the lowest and highest gain along the trajectory. The lower the variation, the more consistent the coverage is. As coVRage operates in UV-space, the results are expressed in UV-coordinates. Small UV-coordinates are nearly equal to the azimuth and elevation in radians.

\subsubsection{Trajectory gain}
First, we analyze the on-trajectory gain. For this, trajectories of similar UV-length (bins of width \SI{0.004}{\uv}\footnote{While UV-coordinates are dimensionless, we introduce the unit \SI{}{\uv} to more clearly indicate (angular) distances in UV-space, with \SI{}{\uv} $\equiv 1$, analogous to the \SI{}{\radian} unit for (dimensionless) radians.}) are grouped together and all sampled gains (at granularity \SI{0.004}{\uv}) are combined per-bin. For visual clarity, the data in the following graphs was smoothed using a Savitzky-Golay filter ($window=11, order=2$)~\cite{savgol}, without affecting any findings. 

Fig.~\ref{fig:trajgood} shows the results for the two beam types, shown using fan charts. In these, lines show median values, while shaded areas cover all samples, with the areas containing the lowest and highest \SI{0.1}{\percent} shaded more darkly. The graph shows that, for trajectories with a length up to \SI{0.5}{\uv}, the tight beam achieves the higher mean gain, while its $0.1^{st}$ percentile is better until \SI{0.3}{\uv}. With longer trajectories, the tight beam begins to degrade, as discussed in Section~\ref{sec:indiv_traj}. This result indicates that the trajectory length for which the optimal beam switches from tight to loose is somewhere between \SI{0.3}{\uv} and \SI{0.5}{\uv}. For short trajectories, the loose beam performs somewhat erratically, as also covered in Section~\ref{sec:indiv_traj}. In this case, a tight beam is advisable. The largest difference in gain between tight and loose beam is \SI{1.775}{\deci\bel}. This means that the tight beam can offer a gain exceeding that of the loose beam by at most \SI{50}{\percent}, in linear terms.

Fig.~\ref{fig:trajbad} evaluates the off-trajectory case, with a shift of \SI{0.075}{\uv}. In such a situation, the loose trajectory performs significantly better, as it achieves its design goal of covering a wider area around the expected trajectory, to account for errors in measurement and prediction. The difference in mean gain is significant for any trajectory length, reaching a maximum of over \SI{9.5}{\dBi}. The tight beam reaches a $0.1^{st}$ percentile of under \SI{-25}{\dBi}. The loose beam again performs somewhat poorly in the lowest percentiles with shorter trajectories. Starting from  \SI{0.5}{\uv} the $0.1^{st}$ percentile does not fall below \SI{-2.5}{\dBi}. As significant measurement and prediction errors are mainly expected for longer trajectories, this is only a limited issue. 

\begin{figure*}[!t]
    \centering
    \begin{minipage}[b][][b]{0.318\textwidth}
        \includegraphics[width=\linewidth]{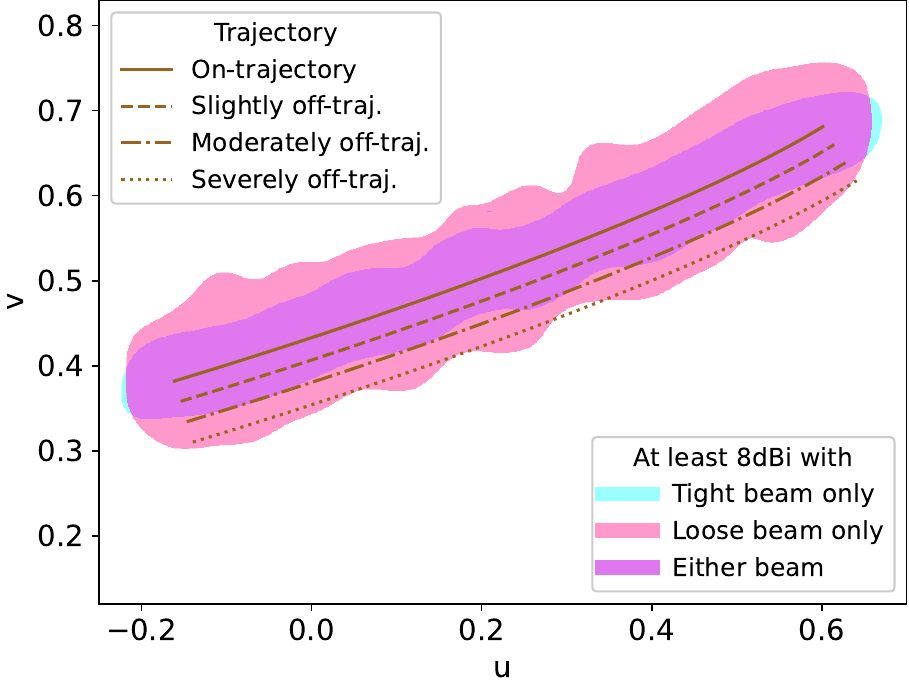}
        \caption{Simplified gain map showing high coverage zones (at least \SI{8}{\dBi}), overlaid with (non-)shifted trajectories}
        \label{fig:trajgainmap}
    \end{minipage}
    \hfill
    \begin{minipage}[b][][b]{0.335\textwidth}
    \includegraphics[width=\linewidth]{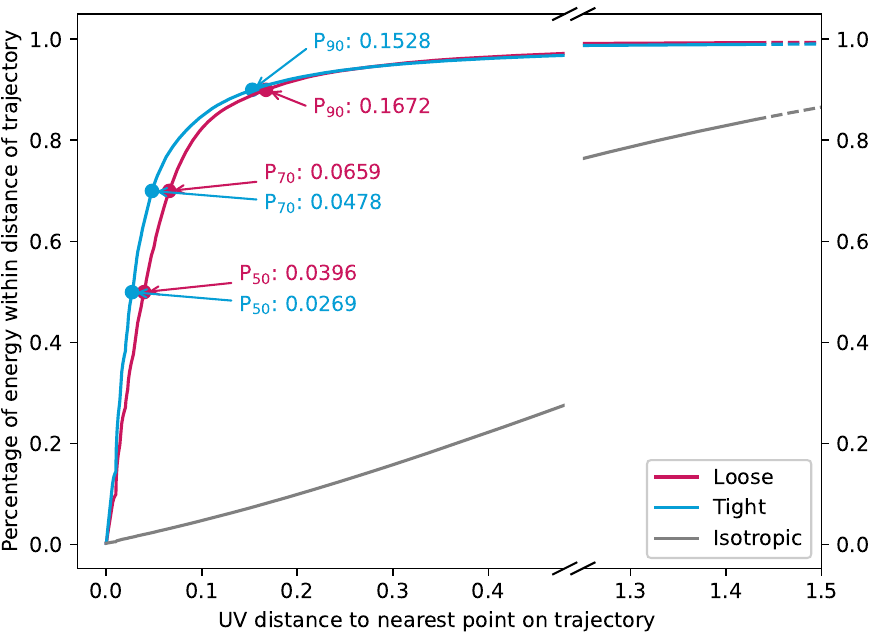}
    \caption{Gain concentration}
    \label{fig:width_cdf}
    \end{minipage}
    \hfill
    \begin{minipage}[b][][b]{0.315\textwidth}
    \includegraphics[width=\linewidth]{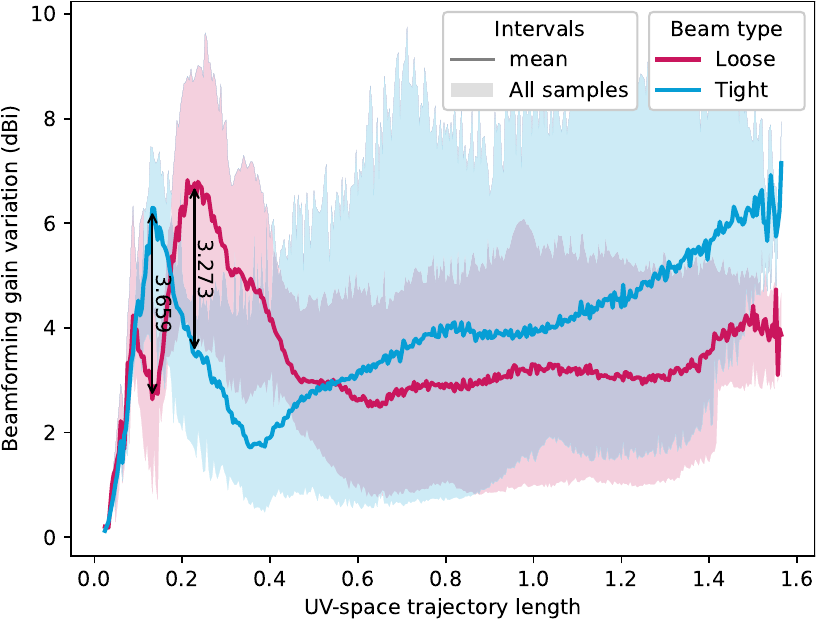}
    \caption{Gain variation}
    \label{fig:trajdist}
    \end{minipage}
\end{figure*}

Finally, Fig.~\ref{fig:trajall} shows the mean gains from the previous two figures, along with less drastically shifted trajectories of \SI{0.025}{\uv} and \SI{0.050}{\uv}, showing that, as intended, the loose beam already outperforms the tight beam even for minor trajectory shifts. Fig.~\ref{fig:trajgainmap} illustrates these shifted trajectories in a simplified gain map for the same beams shown in Fig.~\ref{fig:ex2loose} and \ref{fig:ex2tight}. Clearly, only the loose beam covers the more significant shifts.
\subsubsection{Gain concentration}
Fig.~\ref{fig:width_cdf} shows the gain concentration, averaged over all trajectories, for both loose and tight beams. For comparison, the plot also includes the gain concentration of a theoretical isotropic receive antenna. As an isotropic antenna has a constant gain for all directions, the curve equals the \gls{CDF} of a randomly chosen direction's distance to the trajectory. As the \glspl{CDF} for both loose and tight beams grow significantly more rapidly than this baseline \gls{CDF}, the algorithm clearly achieves its main goal of focusing energy around the trajectory. 

It additionally shows that the tight beam accomplishes its goal of focusing energy more tightly around the trajectory. The largest percentage point difference is achieved in the band containing all directions at most \SI{0.036}{\uv} removed from the trajectory, which contains, on average, \SI{60.5}{\percent} of all energy with a tight beam but only \SI{46.5}{\percent} in the loose case. Conversely, the patch containing half of all energy is \SI{47.2}{\percent} wider for the loose beam compared to the tight beam, as shown with percentile values in the plot. Starting from \SI{94.4}{\percent} of all energy however, the loose beam does overtake the tight beam. We hypothesize that this occurs due to sub-beams from interleaved sub-arrays partially fusing together, effectively reducing the inter-element spacing, which is known to reduce side lobe strength~\cite{phaseBook}. As the tight beam does perform better near the trajectory, this phenomenon does not detract from the usefulness of the tight beam. Note that nonzero energy faraway from the trajectory is entirely unavoidable with phase-shifting arrays, as phases cannot perfectly cancel each other out for a large, continuous range of directions. 

For both beam types, the gain concentration for very short and long trajectories was significantly lower compared to the average case, due to coVRage's reduced effectiveness in such cases, as described above. As the relative reduction in concentration was similar between loose and tight beams, and as these cases are relatively rare, further evaluation of gain concentration as a function of trajectory length would provide no significant additional insight.
\subsubsection{Gain variation}
Fig.~\ref{fig:trajdist} shows the gain variation for the two beam types, as a function of the trajectory length. Both types experience a large peak with short trajectories, with the same cause as previous similar phenomena. Starting from \SI{0.5}{\uv}, the variation for loose beams stabilizes around \SI{3}{\dBi}. This indicates that coVRage achieves a rather even distribution of energy across the trajectory, with the best-covered direction along the trajectory receiving roughly twice as much energy as the worst-covered direction. Ideally, the variation would be zero, however this is again not perfectly achievable with phase shifting. The variation starts to deteriorate slightly for very long trajectories, increasing by around \SI{1}{\dBi}. For tight beams, it steadily deteriorates starting from length 0.4, becoming higher than the loose beam's around \SI{0.55}{\uv}, again showing that tight beams are mainly suitable for shorter trajectories.

Overall, these three metrics together show how coVRage is able to consistently distribute nearly all available energy evenly along the expected trajectory. 
\subsection{Single-block arrays}
\begin{figure*}[t]
    \begin{minipage}{0.64\textwidth}
    \centering
    \subfloat[On-trajectory gain]{\includegraphics[width=0.485\linewidth]{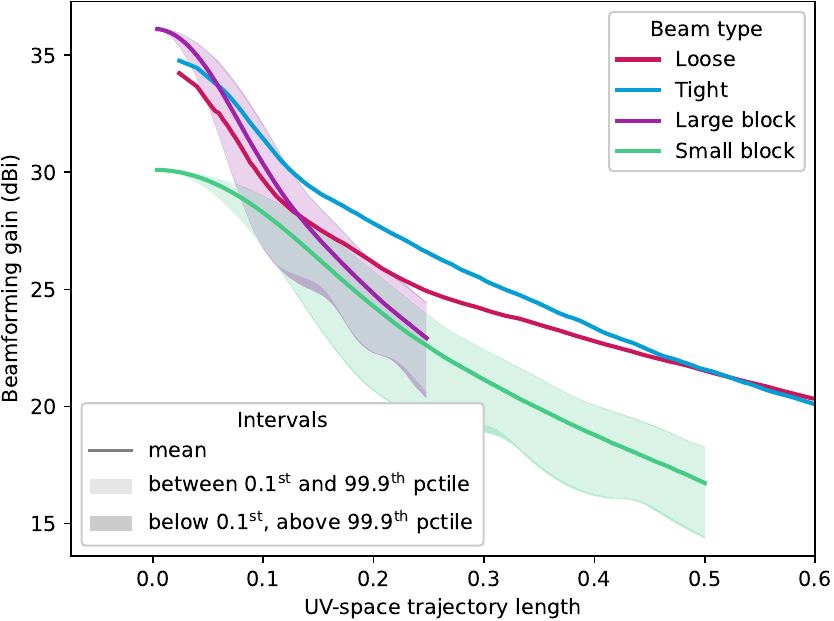} \label{fig:trajgoodbonus}}
    \hfill
    \subfloat[Gain variation]{\includegraphics[width=0.485\linewidth]{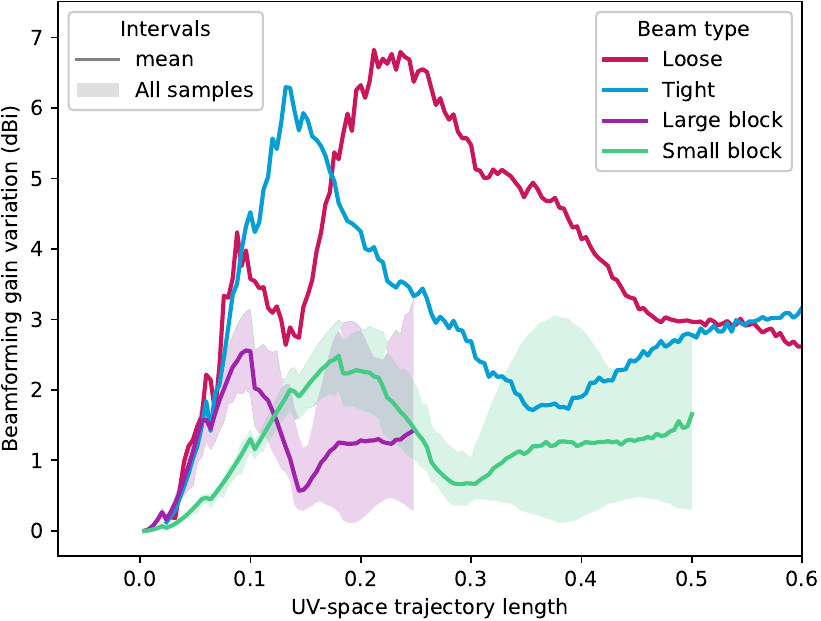} \label{fig:trajdistbonus}}
    \caption{Gain metrics for single-block solutions, with one small ($32 \times 32$) or large ($64 \times 64$) array of four interleaved sub-arrays. The shown UV-range is reduced, and shading for multi-block solutions are removed, to focus on single-block results. Full multi-block results are shown in Fig.~\ref{fig:trajgood} and \ref{fig:trajdist}.}
    \label{fig:trajextra}
    \end{minipage}
    \hfill
    \begin{minipage}{0.325\textwidth}
    \strut\vspace*{-0.3\baselineskip}\newline
    \includegraphics[width=\linewidth]{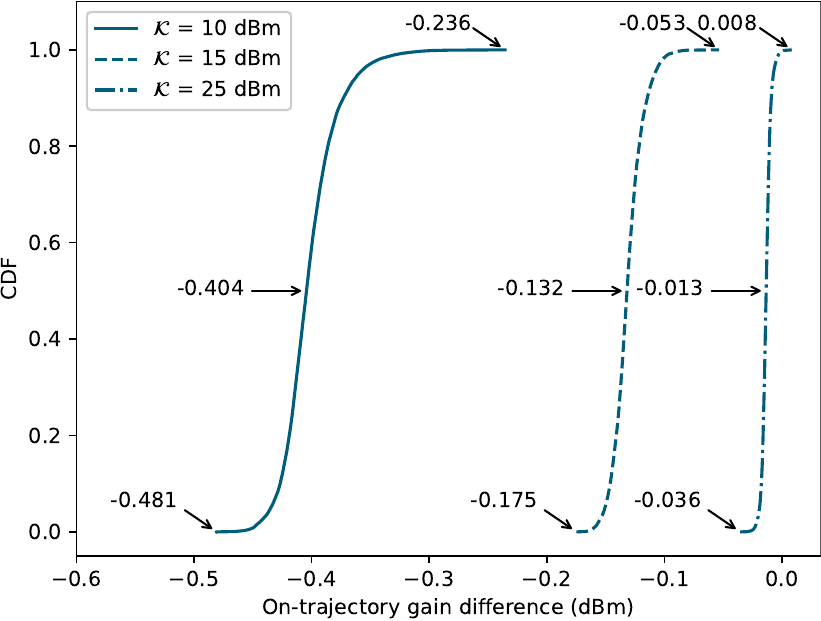}
    \caption{CDF of the change in on-trajectory gain when including 2 \gls{NLoS} paths in the model, averaged over $250$ random initializations of said paths.}
    \label{fig:nlos}
    \end{minipage}
\end{figure*}
\begin{table}[t]
        \caption{Gain diff., quantized vs. continuous (\SI{}{\dB})}
        \centering
        \begin{tabular}{lcccc}
        \toprule
        &\multicolumn{4}{c}{\textbf{Quantization bits}} \\
        & 2 & 4 & 6 & 8 \\
        \cmidrule(lr){2-5}
        $P_{99}$ & 2.68 & 0.480 & 0.115 & 0.0287 \\
        $P_{99.9}$ & 4.95 & 0.758 & 0.181 & 0.0450 \\
        Worst case & 59.9 & 1.65 & 0.496 & 0.0995 \\
        \bottomrule
        \end{tabular}
        \label{tab:quant}
    \end{table}%

The previous results, such as Fig.~\ref{fig:trajdist}, show that there is a range of short trajectories for which the signal strength is inconsistent. As the proximity of sub-beams is responsible for this, reducing the number of sub-beams may alleviate this phenomenon. We consider two methods of achieving this: merging all four blocks into one \textit{large} $64 \times 64$ block, or simply disabling three of the blocks, leaving one \textit{small} $32 \times 32$ block. Both leave four sub-beams, with the former's being twice as tight. For each of the two solutions, we again perform \num{100000} simulations, restricting trajectory length to between \SI{0}{\degree} and \SI{30}{\degree}. In Fig.~\ref{fig:trajextra}, we compare the on-trajectory gain and gain variation of these new results to those previously discussed. Given the lower maximal UV-space trajectory length with only four sub-beams, we restrict the results to \SI{0.25}{\uv} and \SI{0.5}{\uv} for the large and small block, respectively. Except for very short trajectories, under \SI{0.13}{\uv}, both multi-block solutions outperform the large block in terms of average on-trajectory gain, while the small block never reaches the average performance of either multi-block solution. 

In terms of gain variation however, both single-block solutions significantly outperform the multi-block solutions for the shorter paths considered. The two display a similar curve, although the small blocks curve is more stretched horizontally, due to its wider sub-beams. These results also show that the overall gain variation pattern, consisting of an initial peak for short trajectories, followed by a trough and a gradual increase, is inherent to the sub-beam system and cannot be avoided for a single combination of sub-beam width and number. This indicates that there may be merit to an extension of coVRage with dynamic sub-beam width and count, although we leave further investigation of this approach for future work. 
\subsection{Phase Shift Quantization}
In the preceding evaluations, we assume that the phase shifters of the antenna array can continuously achieve any angle with perfect accuracy. In practice, phase shifters only provide a fixed, limited set of angles. An $n$-bit phase shifter supports $2^n$ angles, leading to a granularity of $\frac{2\pi}{2^n}$. To investigate how well coVRage performs with quantized phase shifts, we perform the simulation for \num{10000} randomly generated trajectories, with $2$, $4$, $6$ and $8$-bit phase shifts, along with a continuous baseline. For every possible direction, we compare the beamforming gain of the quantized and continuous systems. To focus on (desired or undesired) high-gain directions only, we filter out any directions where neither system achieves at least \SI{10}{\dBi}, which leaves around five million data points per quantization level. 

Table~\ref{tab:quant} shows the $99^{th}$, $99.9^{th}$ and worst-case absolute difference between the quantized and continuous systems. $2$-bit phase shifters occasionally experience significant gain differences, making them ill-suited for achieving reliably high gain with coVRage. Starting from $4$-bit, the difference is at most \SI{1.65}{\dBi}, with \SI{99}{\percent} under \SI{0.5}{\dBi}. The latter is achieved for all samples with $6$ bits, and an $8$-bit solution further reduces the worst-case difference to under \SI{0.1}{\dBi}. Considering the gain fluctuation throughout the trajectory (Fig.~\ref{fig:trajdist}), coVRage is viable with $4$-bit phase shifters, and near-indistinguishable from the continuous case with $8$-bit phase shifters.
\subsection{Multipath Model}
All preceding experiments use the simplified \gls{LoS}-only model due to runtime concerns. Here, we compare the results of the \gls{LoS}-only and multipath models for a modest set of trajectories to show that the two produce similar results. The simplified model with \gls{NLoS} paths is equivalent to the full model with a single-element isotropic transmitter. Introducing a beamformed transmitter aimed reasonably well towards the receiver would only reduce the impact of \gls{NLoS} paths.

The multipath model features two additional parameters: the number of paths $L$ and the Rician K-factor $\mathcal{K}$, representing the relative strength of the \gls{LoS} path. We set $L$ to 3, in line with related works works, e.g.,~\cite{VirtualHierarchical}, as the mmWave channel is known to be sparse. As \gls{mmWave} K-factor values vary significantly across the literature~\cite{VirtualHierarchical,kfactorMeasure1, kfactorMeasure2, kfactorMeasure3, kfactorModel1}, we perform separate simulations with $\mathcal{K} = $10, 15 and 25 \SI{}{\deci\bel}. For 128 uniformly sampled trajectories, we run $250$ simulations with randomly sampled path coefficients and \glspl{AoA}, both with a loose and a tight beam, for all three K-factors, for a total of \SI{192000}{} simulations. Fig.~\ref{fig:nlos} shows the \gls{CDF} of the gain difference for all points on all considered trajectories. It shows that, on average, the worst gain reduction experienced for any point on any trajectory was under \SI{0.5}{\dBi} for the strongest \gls{NLoS} paths and under \SI{0.04}{\dBi} for the weakest. Compared to the gain fluctuation along a trajectory in Fig.~\ref{fig:trajdist}, the impact of \gls{NLoS} paths on coVRage's performance is negligible.

In the above, we average each result across $250$ simulations. In contrast, Table~\ref{tab:nlos} shows the (near-)worst-case percentiles without averaging, emphasizing outlier results. This shows that, while a significant gain reduction occur very occasionally with lower K-factors, the reduction remains modest for higher K-factors even in the worst case. As the K-factor is known to increase with increasing frequency~\cite{kfactorIncrease}, we expect the higher K-factors to give more accurate results. The negligible impact of the \gls{NLoS} paths and higher runtime of the full model supports our \gls{LoS}-only approach to the full evaluation of coVRage.

\subsection{Practical Limitations}
\begin{table}
    \caption{On-trajectory gain difference with \gls{NLoS} links (\SI{}{\dB})}
    \centering
    \begin{tabular}{lccc}
    \toprule
    &\multicolumn{3}{c}{\textbf{$\mathcal{K}$-factor} (\SI{}{\dB})} \\
    & 10 & 15 &25 \\
    \cmidrule(lr){2-4}
    $P_{99.99}$ & -5.423 & -2.662 & -0.7532 \\
    $P_{99.9999}$ & -11.38 & -5.931 & -1.225 \\
    Worst case & -17.68 & -10.57 & -1.576 \\
    \bottomrule
    \end{tabular}
    \label{tab:nlos}
\end{table}
Beamforming is based on wave interference, i.e. the interaction of waves emitted by the different array elements. Nearby antenna elements will however also interact from an electromagnetic perspective, as they absorb, re-radiate, reflect and scatter each other's signals~\cite{MutualCoupling1,MutualCoupling2}. This process, called mutual coupling, becomes more prominent as inter-element spacing shrinks~\cite{mutualSpacing} and has numerous effects, with a non-negligible (but not necessarily negative) effect on beam gain. As this effect is highly computationally complex to model and dependent on physical characteristics of the specific antenna array, the effect is customarily not considered in information-theoretic evaluations, such as ours. Numerous hardware mitigation techniques against mutual coupling, orthogonal to the beamforming process, have been proposed~\cite{MutualCoupling1,MutualCoupling2}.

\section{Conclusions}\label{sec:conclusion}
In this paper, we presented coVRage, the first beamforming algorithm designed specifically for \gls{HMD}-side beamforming in mobile \gls{VR} applications, where uninterrupted reception even during fast head rotations is crucial for maintaining \gls{QoE}. Using the \gls{HMD}'s built-in orientation detection capabilities, a predictor can estimate how the \gls{AoA} of incoming wireless video data will change in the near future (e.g., next \SI{200}{\milli\second}). By dynamically subdividing the phased array into virtual sub-arrays and aiming each sub-array's beam at a different point along the predicted trajectory, coVRage offers uninterrupted coverage along the full trajectory. At \SI{120}{\giga\hertz}, coVRage can also dynamically change the beamwidth depending on confidence in the measurements and predictions. Through simulation, we evaluated the performance of coVRage for a large range of realistic rotational trajectories. We showed that the algorithm consistently provides a high and stable gain along the trajectory. The dynamic coverage width provides a tunable robustness to temporal and spatial errors in measurements and predictions. Furthermore, we showed that coVRage remains functional with quantized phase shifters and under the influence of reflections. Overall, coVRage provides an important step towards reliable connectivity for truly wireless and highly mobile interactive \gls{VR} with future \glspl{HMD} supporting extremely high resolutions and refresh rates. In future work, we will extend coVRage to the scenario where multiple \glspl{AP} cover the \glspl{HMD}~\cite{vrmultiap}. In addition, we envision an enhancement using Intelligent Reflective Surfaces~\cite{irs}, which could provide \gls{LoS} communication even when such a link is not directly available between \gls{AP} and \gls{HMD}, expected to occur mainly when a user looks up or down.
\section*{Acknowledgment}
The work of Jakob Struye was supported by the Research Foundation - Flanders (FWO): PhD Fellowship 1SB0719N. Filip Lemic acknowledges the in-part support from the MCIN / AEI / 10.13039 / 501100011033 / FEDER / UE HoloMit 2.0 (nr. PID2021- 126551OB-C21). In addition, the work of Filip Lemic at the University of Antwerp - imec was supported by the EU Marie Skłodowska-Curie Actions Individual Fellowships (MSCA IF) project Scalable Localization-enabled In-body Terahertz Nanonetwork (SCaLeITN), grant nr. 893760. This work was supported by the CHIST-ERA grant SAMBAS (CHIST-ERA-20-SICT-003), with funding from FWO, ANR, NKFIH, and UKRI. This research is partially funded by the FWO WaveVR project (Grant number: G034322N). The computational resources and services used in this work were provided by the HPC core facility CalcUA of the Universiteit Antwerpen, and VSC (Flemish Supercomputer Center), funded by FWO and the Flemish Government.
\bibliographystyle{IEEEtran}
\bibliography{IEEEabrv,bibliography}
\end{document}